\documentclass[12pt]{article}
\usepackage{epsfig,cite}
\usepackage{amsmath}
\usepackage{amssymb}
\usepackage{color}
\usepackage{graphicx}
\usepackage{verbatim,mathrsfs,subfigure,feynmf}
\usepackage{url}
\include{epsf}
\newcount\timecount
\newcount\hours \newcount\minutes  \newcount\temp \newcount\pmhours
\hours = \time
\divide\hours by 60
\temp = \hours
\multiply\temp by 60
\minutes = \time
\advance\minutes by -\temp
\def\hour{\the\hours}
\def\minute{\ifnum\minutes<10 0\the\minutes
            \else\the\minutes\fi}
\def\clock{
\ifnum\hours=0 12:\minute\ AM
\else\ifnum\hours<12 \hour:\minute\ AM
      \else\ifnum\hours=12 12:\minute\ PM
            \else\ifnum\hours>12
                 \pmhours=\hours
                 \advance\pmhours by -12
                 \the\pmhours:\minute\ PM
                 \fi
            \fi
      \fi
\fi
}

\def\monthname{\relax\ifcase\month 0/\or January\or February\or
   March\or April\or May\or June\or July\or August\or September\or
   October\or November\or December\else\number\month/\fi}

\def\bold#1{\setbox0=\hbox{$#1$}%
     \kern-.025em\copy0\kern-\wd0
     \kern.05em\copy0\kern-\wd0
     \kern-.025em\raise.0433em\box0 }

\def\beq{\begin{equation}}
\def\eeq{\end{equation}}

\def\ga{\mathrel{\raise.3ex\hbox{$>$\kern-.75em\lower1ex\hbox{$\sim$}}}}
\def\la{\mathrel{\raise.3ex\hbox{$<$\kern-.75em\lower1ex\hbox{$\sim$}}}}
\def\gev{{\rm \, Ge\kern-0.125em V}}
\def\tev{{\rm \, Te\kern-0.125em V}}
\def\gyr{{\rm \, G\kern-0.125em yr}}

\def\gappeq{\mathrel{\rlap {\raise.5ex\hbox{$>$}}
{\lower.5ex\hbox{$\sim$}}}}
\def\lappeq{\mathrel{\rlap{\raise.5ex\hbox{$<$}}
{\lower.5ex\hbox{$\sim$}}}}
\def\Toprel#1\over#2{\mathrel{\mathop{#2}\limits^{#1}}}

\newcommand{\vphi}{\varphi}

\def\m12{m_{1\!/2}}

\def\bea{\begin{eqnarray}}
\def\eea{\end{eqnarray}}

\usepackage{lscape}

\def\beq{\begin{equation}}
\def\eeq{\end{equation}}

\def\mgut{M_{GUT}}

\begin{document}

\begin{titlepage}
\pagestyle{empty}
{\tt
\rightline{KCL-PH-TH/2017-05, CERN-TH/2017-010}
\rightline{KIAS-P17008, UT-17-03}
\rightline{ACT-01-17; MI-TH-1738}
\rightline{UMN-TH-3618/17, FTPI-MINN-17/02}
}
\vspace{0.8cm}
\begin{center}
{\bf {\Large No-Scale SU(5) Super-GUTs} }
\end{center}

\vspace{0.25cm}
\begin{center}
{\bf John~Ellis}$^{1,2}$,
{\bf Jason L. Evans}$^3$,
{\bf Natsumi Nagata}$^{4}$, \\
\vskip 0.1in
{\bf Dimitri V. Nanopoulos}$^{5,6,7}$
and {\bf Keith~A.~Olive}$^{8}$
\vskip 0.2in
{\small {\it
$^1${Theoretical Physics and Cosmology Group, Department of Physics, \\ King's College London, Strand,
London~WC2R 2LS, UK}\\
\vspace{0.2cm}
$^2${Theoretical Physics Department, CERN, CH-1211 Geneva 23, Switzerland}\\
\vspace{0.2cm}
$^3${School of Physics, KIAS, Seoul 130-722, Korea}\\
\vspace{0.2cm}
$^4${Department of Physics, University of Tokyo, Bunkyo-ku, Tokyo 113--0033, Japan}\\
\vspace{0.2cm}
$^5${George P. and Cynthia W. Mitchell Institute for Fundamental Physics and Astronomy,
Texas A\&M University, College Station, TX 77843, USA}\\
\vspace{0.2cm}
$^6${Astroparticle Physics Group, Houston Advanced Research Center (HARC), \\ Mitchell Campus, Woodlands, TX 77381, USA}\\
\vspace{0.2cm}
$^7${Academy of Athens, Division of Natural Sciences,
Athens 10679, Greece}\\
\vspace{0.2cm}
$^8${William I. Fine Theoretical Physics Institute, School of Physics and Astronomy,\\
University of Minnesota, Minneapolis, MN 55455,\,USA}}}\\

\vspace{1cm}
{\bf Abstract}
\end{center}
{\small
 We reconsider the minimal SU(5) Grand Unified Theory (GUT) in the
 context of no-scale supergravity, assuming that the soft
 supersymmetry-breaking parameters satisfy universality conditions at
 some input scale $M_{\rm in}$ above the GUT scale $\mgut$. When setting
 up such a no-scale super-GUT model, special attention must be paid to
 avoiding the Scylla of rapid proton decay and the Charybdis of an
 excessive density of cold dark matter, while also having an acceptable
 mass for the Higgs boson. We do not find consistent solutions if none
 of the matter and Higgs fields are assigned to twisted chiral
 supermultiplets, even in the presence of Giudice-Masiero
 terms. However, consistent solutions may be found if at least one
 fiveplet of GUT Higgs fields is assigned to a twisted chiral
 supermultiplet, with a suitable choice of modular weights.
 Spin-independent dark matter scattering may be detectable
 in some of these consistent solutions.}


\vfill
\leftline{February 2017}
\end{titlepage}

\section{Introduction}
\label{sec:intro}

Globally supersymmetric grand unification has long been an attractive
framework for unifying the non-gravitational interactions, with the
minimal option using the gauge group SU(5)
\cite{GG,Dimopoulos:1981zb}. When incorporating gravity, one must embed
such a supersymmetric Grand Unified Theory (GUT) within some
supergravity theory, and an attractive option is no-scale
supergravity~\cite{noscale}. This has the advantages that it leads to an
effective potential without holes of depth ${\cal O}(1)$ in natural
units, and emerges in generic string
compactifications~\cite{Witten}. No-scale supergravity also allows
naturally for the possibility of Planck-compatible cosmological
inflation~\cite{EGNO}. In general, a no-scale K\"ahler potential
contains several moduli $T_i$, but here we consider scenarios in which
the relevant dynamics is dominated by a single volume modulus field $T$.

The construction of no-scale supergravity GUTs encounters
significant hurdles, such as fixing the compactification moduli. Moreover,
pure no-scale boundary conditions require that all the
quadratic, bilinear and trilinear scalar couplings $m_0, B_0$ and $A_0$ vanish,
leading to phenomenology that is in contradiction with experimental constraints.
However, this issue may be avoided in models with (untwisted or twisted)
matter fields with non-vanishing modular weights as we show below.

The simplest possibility for soft supersymmetry breaking is to postulate
universal values of $m_0, B_0$ and $A_0$, as in the constrained minimal
supersymmetric Standard Model (CMSSM)~\cite{funnel,cmssm,elos,eelnos}.
With the inclusion of a universal gaugino mass, $m_{1/2}$, the CMSSM is
a four-parameter theory\footnote{In addition, one must choose the sign
of $\mu$ which we take here to positive.}. Minimal supergravity places
an additional boundary condition, relating $B_0$ and $A_0$ ($B_0 = A_0 -
m_0$) making it a three-parameter theory \cite{bfs,vcmssm}. No-scale
supergravity, however is  effectively a one-parameter theory since we
require $m_0 = A_0 = B_0 = 0$. Another one-parameter theory in this
context is pure gravity mediation \cite{pgm,eioy,eioy2}, in which the
gaugino masses, $A$ and $B$ terms\footnote{In order to get electroweak symmetry breaking to work, the $B$
terms in these models also get a contribution from a Giudice-Masiero term~\cite{gm}.} are determined by anomaly mediation
\cite{anom} leaving only the gravitino mass, $m_{3/2} = m_0$ as a free
parameter.

These boundary conditions may be too restrictive if they are imposed at
the GUT scale, $\mgut$, defined as the renormalization scale where the
two electroweak gauge couplings are unified.  There is, however, no
intrinsic reason that the boundary conditions for supersymmetry breaking
coincide with gauge coupling unification. Separating these two scales
opens the door for so-called sub-GUT models \cite{subGUT,elos, eelnos}
where the input universality scale differs from the GUT scale with
$M_{in} < \mgut$ or the possibility that the boundary conditions are
imposed at some higher input scale $M_{\rm in} > \mgut$, a scenario we
term super-GUT~\cite{emo, eemno}.

However, the regions of parameter space with acceptable relic density and
Higgs mass typically require quite special values of the GUT superpotential
couplings and rather large values of $\tan \beta$ \cite{emo2}, and hence a
proton lifetime that is unacceptably short. In order to accommodate smaller
values of $\tan \beta$ and hence an acceptably long proton lifetime, we
consider non-zero Giudice-Masiero (GM) terms~\cite{gm} in the K\"ahler
potential. In this way we are able to avoid the Scylla of rapid proton
decay and the Charybdis of an excessive density of cold dark matter,
while also having an acceptable value of the Higgs mass. Furthermore, when
no-scale boundary conditions are applied at the GUT scale, the lightest sparticle  in the spectrum
is typically a stau (or the stau is tachyonic). Applying the boundary conditions
above the GUT scale as in a super-GUT model can alleviate this problem \cite{eno5}.

The outline of this paper is as follows. In Section~\ref{sec:super-GUT},
we review our theoretical framework, with our set-up of the minimal
supersymmetric SU(5) model described in Subsection~\ref{dec:msgut}, our
no-scale supergravity framework described in
Subsection~\ref{sec:noscale} and the vacuum conditions and the relevant
renormalization-group equations (RGEs) set out in
Subsection~\ref{sec:vaccond}. We describe our key results in
Section~\ref{sec:results}. We explore in Section~\ref{sec:gmterm}
scenarios in which none of the matter and Higgs supermultiplets are
twisted, and find no way to steer between Scylla and Charybdis with an
acceptable Higgs mass in this case. However, as we show in
Section~\ref{sec:twisted}, this is quite possible if one or the other
(or both) of the GUT fiveplet Higgs supermultiplets is twisted. Spin-independent
dark matter scattering may be observable in some of the cases studied. Finally,
Section~\ref{sec:discussion} discusses our results.

\section{Super-GUT CMSSM Models}
\label{sec:super-GUT}

\subsection{Minimal Supersymmetric  SU(5)}
\label{dec:msgut}

The minimal supersymmetric SU(5) GUT~\cite{Dimopoulos:1981zb},
was recently reviewed in \cite{eemno} and we recall here the aspects most needed
for our discussion.
The minimal renormalizable superpotential for this model is
given by
\begin{align}
 W_5 &=  \mu_\Sigma {\rm Tr}\Sigma^2 + \frac{1}{6} \lambda^\prime {\rm
 Tr} \Sigma^3 + \mu_H \overline{H} H + \lambda \overline{H} \Sigma H
\nonumber \\
&+ \left(h_{\bf 10}\right)_{ij} \epsilon_{\alpha\beta\gamma\delta\zeta}
 \Psi_i^{\alpha\beta} \Psi^{\gamma\delta}_j H^\zeta +
 \left(h_{\overline{\bf 5}}\right)_{ij} \Psi_i^{\alpha\beta} \Phi_{j \alpha}
 \overline{H}_\beta ~,
\label{W5}
\end{align}
where Greek sub- and superscripts denote SU(5) indices, and $\epsilon$
is the totally antisymmetric tensor with $\epsilon_{12345}=1$. In
Eq.~\eqref{W5}, the adjoint multiplet $\Sigma \equiv \sqrt{2}\Sigma^A
T^A$, where the $T^A$ ($A=1, \dots, 24$) are the generators of SU(5)
normalized so that ${\rm Tr}(T^A T^B) = \delta_{AB}/2$, is responsible
for breaking SU(5) to the Standard Model (SM).
The scalar components of $\Sigma$ are assumed to have vevs of the form
\begin{equation}
 \langle \Sigma \rangle = V \cdot {\rm diag} \left(2,2,2,-3,-3\right) ~,
\end{equation}
where $V\equiv 4 \mu_\Sigma/\lambda^\prime$,
causing the GUT gauge bosons $X$ to acquire masses $M_X = 5 g_5 V$,
where $g_5$ is the SU(5) gauge coupling.

The multiplets $H$ and $\overline{H}$ in Eq.~\eqref{W5} are  ${\bf 5}$ and $\overline{\bf 5}$
representations of SU(5), respectively, and contain the MSSM Higgs fields.
In order to realize
doublet-triplet mass splitting in the $H$ and $\overline{H}$ multiplets, we
impose the fine-tuning condition $\mu_H -3\lambda V \ll V$.
In this case, the color-triplet Higgs states have masses $M_{H_C} = 5\lambda V$,
the masses of the
color and weak adjoint components of $\Sigma$ are $M_\Sigma =
5\lambda^\prime V/2$, and the singlet component of $\Sigma$ acquires a
mass $M_{\Sigma_{24}} = \lambda^\prime V/2$.

The multiplets ${\Phi}_i$ in Eq.~\eqref{W5} are  $\bf{\overline{5}}$ representations containing the left-handed SM matter fields
$\overline{D}_i$ and ${L}_i$, and the ${\Psi}_i$ are $\bf{10}$ representations of SU(5) containing the left-handed
${Q}_i$, $\overline{U}_i$, and $\overline{E}_i$,
where the index $i = 1,2,3$ denotes the
generations.

The soft supersymmetry-breaking terms in the minimal supersymmetric SU(5) GUT are
\begin{align}
 {\cal L}_{\rm soft} = &- \left(m_{\bf 10}^2\right)_{ij}
 \widetilde{\psi}_i^* \widetilde{\psi}_j
- \left(m_{\overline{\bf 5}}^2\right)_{ij} \widetilde{\phi}^*_i
 \widetilde{\phi}_j
- m_H^2 |H|^2 -m_{\overline{H}}^2 |\overline{H}|^2 - m_\Sigma^2 {\rm Tr}
\left(\Sigma^\dagger \Sigma\right)
\nonumber \\
&-\biggl[
\frac{1}{2}M_5 \widetilde{\lambda}^{A} \widetilde{\lambda}^A
+ A_{\bf 10} \left(h_{\bf 10}\right)_{ij}
 \epsilon_{\alpha\beta\gamma\delta\zeta} \widetilde{\psi}_i^{\alpha\beta}
 \widetilde{\psi}^{\gamma\delta}_j H^\zeta
+ A_{\overline{\bf 5}}\left(h_{\overline{\bf 5}}\right)_{ij}
 \widetilde{\psi}_i^{\alpha\beta} \widetilde{\phi}_{j \alpha}  \overline{H}_\beta
\nonumber \\
&+ B_\Sigma \mu_\Sigma {\rm Tr} \Sigma^2 +\frac{1}{6} A_{\lambda^\prime
 } \lambda^\prime  {\rm Tr} \Sigma^3 +B_H \mu_H \overline{H} H+
 A_\lambda \lambda \overline{H} \Sigma H +{\rm h.c.}
 \biggr]~,
\end{align}
where $\widetilde{\psi}_i$ and $\widetilde{\phi}_i$ are the scalar
components of $\Psi_i$ and $\Phi_i$, respectively,
the $\widetilde{\lambda}^A$ are the SU(5) gauginos. We
use the same symbols for the scalar components of the Higgs fields as for the
corresponding superfields.

\subsection{No-Scale Framework}
\label{sec:noscale}

We refer to~\cite{egno4} for a derivation of the soft terms arising in no-scale
supergravity~\footnote{Related derivations of soft terms in string models with flux compactifications can be found in \cite{lnr}.}.
Our starting-point is a no-scale K\"ahler potential
\beq
K \; = \; - 3 \ln \left(T + {\bar T} - \frac{1}{3} \sum_i |\phi_i|^2\right) + \sum_a \frac{|\vphi_a|^2}{(T + {\bar T})^{n_a}} \, ,
\label{finalK}
\eeq
which includes a volume modulus field, $T$, and both untwisted and twisted matter fields, $\phi_i$
and $\vphi_a$ respectively, the latter with modular weights $n_a$.
We consider a generic superpotential of the form
\beq\label{w_phi}
\begin{aligned}
W&= (T+c)^{\beta}W_2(\phi_i) + (T+c)^{\alpha}W_3(\phi_i) \\
&\qquad +(T+c)^{\sigma}W_2(\vphi_a) +(T+c)^{\rho}W_3(\vphi_a) +  \mu_\Lambda \, ,
\end{aligned}
\eeq
where $c$ is an arbitrary constant, and
$W_{2,3}$ denote bilinear and trilinear terms with modular weights that are in general non-zero.
When $\langle \phi,\vphi \rangle=0$,
the effective potential for $T$ is completely flat at the tree level, so it has an undetermined vev, and
the gravitino mass
\beq
m_{3/2} = \frac{\mu_\Lambda}{(T+\bar{T})^{3/2}}
\eeq
varies with the value of this volume modulus.
We assume here that some Planck scale dynamics fixes $T = \bar{T} = c$, and take $c = 1/2$ in the following.

In a standard no-scale supergravity model with no twisted fields and with weights $\alpha = \beta = 0$,
we would obtain $m_0 = A_0 = B_0  = 0$. However, in the scenario (\ref{w_phi})
soft terms are induced, as were calculated in \cite{egno4}, which
are sector-dependent:
\begin{align}\label{soft_T_gen}
\phi_i: & \quad m_0=0\ , \quad B_0 = -\beta m_{3/2} \ , \quad
A_0=-\alpha m_{3/2} \, ,
\end{align}
\begin{align}\label{soft_T_genvphi}
\vphi_a: & \quad
m_0=m_{3/2}\ , \quad
B_0 = 2m_{3/2} \left(1-\frac{\sigma}{2}\right) \ , \qquad
A_0 = 3m_{3/2} \left(1-\frac{\rho}{3}\right) \, ,
\end{align}
where we have assumed for simplicity that $n_a = 0$.
We also postulate in what follows generalized Giudice-Masiero terms \cite{gm}
\beq\label{GM1}
\Delta K = \left(c_H (T+c)^{\gamma_H} H \bar{H} + c_\Sigma (T+c)^{\gamma_\Sigma} \Sigma^2 + {\rm h.c.}\right) .
\eeq
If $H$, $\bar{H}$, and $\Sigma$ are untwisted, these induce corrections
to the $\mu$ and $B$ terms:
\beq
\Delta \mu_{H} = c_H m_{3/2}\ ,~~ \Delta \mu_{\Sigma} = c_\Sigma m_{3/2}\ , ~~
\Delta B_H \mu_H = -\gamma_H c_H m_{3/2}^2\ , ~~
\Delta B_\Sigma \mu_\Sigma = -\gamma_\Sigma c_\Sigma m_{3/2}^2 .
\eeq
If the fields are twisted, the shift in the $\mu$-terms is the same, but
the shift in $B \mu$ is modified by $-\gamma_{H,\Sigma} c_{H,\Sigma} \to (2-\gamma_{H,\Sigma}) c_{H,\Sigma}$ \cite{egno4}.  Although the corrections to the $B$
terms are quite small, they are crucial for matching the GUT scale $B$ terms onto the MSSM $B$ term at the GUT scale,
as we see below.

In the super-GUT version of the CMSSM model we impose the following universality
conditions for the soft mass parameters at a soft supersymmetry-breaking
mass input scale
$M_{in} > M_{\rm GUT}$:
\begin{align}
 \left(m_{\bf 10}^2\right)_{ij} =
\left(m_{\overline{\bf 5}}^2\right)_{ij}
&\equiv m_0^2 \, \delta_{ij} ~,
\nonumber \\[3pt]
m_H = m_{\overline{H}} = m_\Sigma &\equiv m_0 ~,
\nonumber \\[3pt]
A_{\bf 10} = A_{\overline{\bf 5}} = A_\lambda = A_{\lambda^\prime}
&\equiv A_0 ~,
\nonumber \\[3pt]
B_H = B_\Sigma &\equiv B_0 ~,
\nonumber \\[3pt]
 M_5 &\equiv m_{1/2} ~,
\label{eq:inputcond}
\end{align}
with the input soft terms $m_0, A_0$ and $B_0$ specified above. In the above expressions, and in expressions throughout the text, the $\Delta B$ contribution is neglected since it is so small. However, this contribution to the $B$-terms is included in all calculations in order to satisfy the $B$-term matching condition.

\subsection{Vacuum Conditions and Renormalization-Group Equations}
\label{sec:vaccond}

Since the $B$-term boundary conditions are specified at $M_{in}$,
we cannot use the Higgs minimization equations to determine $B$ and the
MSSM $\mu$ term as is commonly done in the MSSM. Instead, as in mSUGRA models,
these conditions can be used to determine $\mu$ and $\tan \beta$ \cite{vcmssm}
as was done in the no-scale super-GUT models considered in \cite{emo2}.
In \cite{emo2}, standard no-scale boundary conditions were used to
identify regions of parameter space with acceptable relic density and
Higgs mass. Typically, rather large values of $\tan \beta$ were found
and, in addition, it was necessary to choose somewhat small values of
the coupling $\lambda =  {\cal O}(0.01)$ with much larger values of
$\lambda^\prime = {\cal O}(1)$. All of these choices
tend to decrease the proton lifetime to unacceptably small values \cite{eemno}.
In order to reconcile the proton lifetime with the relic density and Higgs mass,
we need to consider lower values of $\tan \beta$ \cite{evno,eelnos,eemno},
which can be accomplished
when the GM terms (\ref{GM1}) are included \cite{gm,dlmmo,eioy,eioy2}.

The soft supersymmetry breaking parameters are evolved down from $M_{in}$ to $M_{GUT}$ using the
renormalization-group equations (RGEs) of the minimal supersymmetric SU(5) GUT,
which can be found in~\cite{pp,Baer:2000gf,emo,emo2,emo3}, with appropriate changes
of notation. During the evolution, the GUT couplings in Eq.~\eqref{W5}
affect the running of the soft supersymmetry-breaking parameters, which results
in non-universality in the soft parameters at $M_{GUT}$. In particular,
the GUT coupling $\lambda$ contributes
to the running of the Yukawa couplings, the corresponding $A$-terms, and the Higgs soft masses.
On the other hand,
$\lambda^\prime$ affects directly only the running of $\lambda$,
$m_\Sigma$, and $A_\lambda$ (besides $\lambda^\prime$ and
$A_{\lambda^\prime}$), and thus can affect the MSSM soft mass parameters
only at higher-loop level. Both $\lambda$ and
$\lambda^\prime $ contribute to the RGEs of the soft masses of
matter multiplets only at higher-loop level, suppressing their effects on
these parameters.

At the unification scale $M_{GUT}$ (defined as the renormalization scale
where the two electroweak gauge couplings are equal), the SU(5) GUT
parameters are matched onto the MSSM parameters. The matching conditions
for the Standard Model gauge and Yukawa couplings were discussed in
detail in \cite{eemno}. The use of threshold corrections at the GUT
scale \cite{Hisano:1992mh, Hisano:1992jj, Hisano:2013cqa}
allow us to determine the SU(5) gauge coupling, $g_5$, and the SU(5)
Higgs adjoint vev, $V$, which in turn allows us to fix the gauge and
Higgs boson masses as
\begin{align}
 M_{H_C} &= 5\lambda V ~,\label{eq:MHCV} \\
 M_\Sigma &= \frac{5}{2} \lambda^\prime V ~, \\
 M_X &= 5 g_5 V ~,\label{eq:MXV}
\end{align}
which are inputs in the calculation of the proton lifetime.

As explained in \cite{eemno}, in order
to allow both $\lambda$ and $\lambda^\prime$ to remain as free
parameters, we must include a Planck-suppressed operator such as
\begin{equation}
W_{\rm eff}^{\Delta g} = \frac{c_5}{M_P} {\rm Tr}\left[ \Sigma {\cal W}
{\cal W}\right] ~,
\label{eq:weffdim5}
\end{equation}
where ${\cal W}\equiv T^A {\cal W}^A$ denotes the
superfields corresponding to the field strengths of the SU(5) gauge
vector bosons. Such operators may make contributions comparable to other
threshold corrections when $\Sigma$ develops a vev  \cite{Ellis:1985jn,
Hill:1983xh, Tobe:2003yj}. We have checked that the coefficient $c_5$ takes
reasonable values, {\it i.e.}, $|c_5| < {\cal O} (1)$.

The matching conditions for the soft supersymmetry-breaking terms were
also discussed in detail in \cite{eemno}. The matching conditions for
the gaugino masses \cite{Tobe:2003yj, Hisano:1993zu} are given by
\begin{align}
 M_1 &= \frac{g_1^2}{g_5^2} M_5
-\frac{g_1^2}{16\pi^2}\left[10 M_5 +10(A_{\lambda^\prime} -B_\Sigma)
 +\frac{2}{5}B_H\right]
+\frac{4c_5 g_1^2V(A_{\lambda^\prime} -B_\Sigma)}{M_P} ~,
\label{eq:m1match}
\\[3pt]
M_2 &= \frac{g_2^2}{g_5^2} M_5
-\frac{g_2^2}{16\pi^2}\left[6 M_5 +6A_{\lambda^\prime} -4B_\Sigma
 \right]
+\frac{12c_5 g_2^2V(A_{\lambda^\prime} -B_\Sigma)}{M_P} ~,
\label{eq:m2match}
\\[3pt]
M_3 &= \frac{g_3^2}{g_5^2} M_5
-\frac{g_3^2}{16\pi^2}\left[4 M_5 +4A_{\lambda^\prime} -B_\Sigma
+B_H \right]
-\frac{8c_5 g_3^2V(A_{\lambda^\prime} -B_\Sigma)}{M_P}
~.
\label{eq:m3match}
\end{align}
We again find that the contribution of the dimension-five operator
in Eq.~\eqref{eq:weffdim5} can be comparable to that of the one-loop
threshold corrections. MSSM soft masses
and the $A$-terms of the third generation sfermions, are given by
\begin{align}
 m^2_{Q} = m_{U}^2 = m^2_{E} = m^2_{{\bf 10}} ~,&
~~~~~~ m_{D}^2 = m_{L}^2 = m_{\overline{\bf 5}}^2 ~, \nonumber \\
m_2 \equiv m_{H_u}^2 = m_H^2 ~,& ~~~~~~ m_1 \equiv m_{H_d}^2 = m_{\overline{H}}^2 ~,
\nonumber \\
 A_t = A_{\bf 10} ~,& ~~~~~~
 A_b = A_\tau = A_{\overline{\bf 5}} ~.
\end{align}
The MSSM $\mu$ and $B$ terms are \cite{Borzumati:2009hu}
\begin{align}
 \mu &= \mu_H - 3 \lambda V\left[
1+ \frac{A_{\lambda^\prime} -B_\Sigma}{2 \mu_\Sigma}
\right] ~,
\label{eq:matchingmu}
 \\[3pt]
 B &= B_H + \frac{3\lambda V \Delta}{\mu}
+ \frac{6 \lambda}{\lambda^\prime \mu} \left[
(A_{\lambda^\prime} -B_\Sigma) (2 B_\Sigma -A_{\lambda^\prime}
 +\Delta) -m_\Sigma^2
\right]~,
\label{eq:matchingb}
\end{align}
with
\begin{equation}
 \Delta \equiv A_{\lambda^\prime} - B_\Sigma - A_\lambda +B_H ~.
\label{eq:deltadef}
\end{equation}
The amount of fine-tuning required to obtain values of $\mu$ and $B$
that are ${\cal O}(M_{\rm SUSY})$ is determined by these last two
equations. From Eq.~\eqref{eq:matchingmu}, we find that we need to tune
$|\mu_H -3\lambda V|$ to be ${\cal O}(M_{\rm SUSY})$. From
Eq.~\eqref{eq:matchingb}, $V\Delta/\mu$ should be ${\cal O}(M_{\rm
SUSY})$, which requires $|\Delta| \leq {\cal O}(M_{\rm
SUSY}^2/M_{GUT})$. In standard no-scale supergravity, $\Delta = 0$ and
this is stable against radiative corrections, as shown in
Ref.~\cite{Kawamura:1994ys}. As discussed in Ref.~\cite{eemno}, in order
for Eq.~\eqref{eq:matchingb} to have a real solution for $B_\Sigma$, the
condition $A_{\lambda^\prime}^2 \gtrsim 8m_\Sigma^2$ should be satisfied
for $\lambda^\prime \ll \lambda$. We have checked that this condition is
always satisfied over the parameter space we consider in
Section~\ref{sec:results}.

The MSSM  $\mu$ and $B$ parameters can be determined by using the electroweak
vacuum conditions:
\begin{align}
 \mu^2 &= \frac{m_{1}^2 -m_{2}^2 \tan^2\beta + \frac{1}{2} m_Z^2
 (1-\tan^2\beta ) +\Delta_\mu^{(1)}}{\tan^2 \beta -1 +\Delta_\mu^{(2)}}
 , \\
 B\mu &= -\frac{1}{2}(m_1^2 + m_2^2 +2 \mu^2) \sin 2\beta +\Delta_B ~, \label{mssmbmu}
\end{align}
where $\Delta_B$ and $\Delta_\mu^{(1,2)}$ denote loop corrections
\cite{Barger:1993gh}.  These are run up to the GUT scale where the
conditions (\ref{eq:matchingmu}) and (\ref{eq:matchingb}) are applied.
However, in standard no-scale supergravity,
the right-hand side of (\ref{eq:matchingb}) is determined by
running down the $A$ and $B$-terms set by $A_0 = B_0 = 0$ (and similarly
for $m_\Sigma^2$). Thus, (\ref{eq:matchingb}) is not satisfied in general.
Nevertheless, it is often possible to find a value of $\tan \beta$ that
adjusts $B\mu$ via  (\ref{mssmbmu}) to have the correct value at the GUT scale.
As noted earlier, this often leads to relatively large values of $\tan
\beta$ and unacceptable low values for the proton lifetime.

Alternatively, we can introduce a GM term in the K\"ahler potential
as in Eq.~\eqref{GM1}. For now, we assume that all fields are untwisted with weight $\gamma = -1$.
The shift in the $\mu$-terms is ${\cal O}(M_{\rm SUSY})$ and is
irrelevant to the matching condition (\ref{eq:matchingmu}). Similarly the shifts in most of the terms in (\ref{eq:matchingb})
are of order $m_{3/2}^2/M_{GUT}$ and are much smaller than ${\cal O}(M_{\rm SUSY})$.
However, there is a shift in $\Delta$
\beq
\delta \Delta = \left(\frac{c_H}{\mu_H} - \frac{c_\Sigma}{\mu_\Sigma}\right) m_{3/2}^2 .
\eeq
Although this shift is also small, $\Delta$ is multiplied by $V/\mu$ in (\ref{eq:matchingb}),
so that the overall shift in $B$ is ${\cal O}(M_{\rm SUSY})$.
Thus the shift in (\ref{eq:matchingb}) becomes
\beq
\frac{3 \lambda V \Delta}{\mu} \to \left(c_H - \frac{12 \lambda}{\lambda^\prime}c_\Sigma \right)
\frac{m_{3/2}^2}{\mu} ,
\label{shift}
\eeq
up to ${\cal O}(M_{\rm SUSY}/M_{\rm GUT})$ corrections. This is now of
comparable size to other terms in (\ref{eq:matchingb}), which can be
satisfied for any $\tan \beta$.  The matching condition
(\ref{eq:matchingb}), therefore determines a linear combination of the
two GM terms.

Our no-scale super-GUT model is therefore
specified by the following set of input parameters:
\begin{equation}
 m_{1/2}, \ M_{in},\ \lambda,\ \lambda',\ \tan \beta,\ {\rm
  sign}(\mu) \, ,
\end{equation}
where the trilinear superpotential Higgs couplings, $ \lambda$ and $\lambda'$,
are specified at $Q=\mgut$.

In the following we assume initially that all fields are untwisted, so that $m_0 = 0$,
and assume vanishing modular weights $\alpha = \beta = 0$, so that $A_0 = B_0 = 0$.
Later we consider the effects of twisting one or both of the Higgs 5-plets
and turning on the trilinear weight $\alpha$ in order to allow non-zero $A_0$.

\section{Results}
\label{sec:results}

\subsection{Standard No-scale Supergravity with a GM Term}
\label{sec:gmterm}

It is well known that the CMSSM with no-scale boundary conditions is not viable.
With $m_0 = A_0 = B_0 = 0$, the particle spectrum almost inevitably contains either
a stau lightest supersymmetric particle (LSP) or tachyonic stau. However, this problem can be alleviated if
the universal boundary conditions are applied above the GUT scale \cite{eno5}.
In this case, the running from $M_{\rm in}$ to $M_{\rm GUT}$ produces non-zero
soft terms that may be sufficiently large to produce a reasonable
spectrum~\footnote{Similar conclusions were reached in gaugino-mediated
models in~\cite{SS}.}.

The basic no-scale super-GUT model was studied in detail in \cite{emo2}.
There it was found that, for sufficiently large $M_{\rm in}$, not only
could a reasonable mass spectrum be obtained, but also
regions of parameter space with the correct relic density and
Higgs mass were identified.  This region was further explored in
\cite{eovs}, with the aim of studying possible departures from minimal
flavor violation. There, for example, a particular benchmark point was chosen
with $M_5 = 1500$~GeV, $M_{\rm in} = 10^{18}$~GeV, $\lambda = -0.1,
\lambda^\prime = 2$, which required $\tan \beta \approx 52$ as no GM
term was included. One concern for this benchmark is the proton decay
rate that is enhanced by the combination of large $\tan \beta$ and
small $\lambda$ (which induced a low value for the Higgs color
triplet mass). Indeed, as we show below, the proton lifetime is far too small
in this minimal SU(5) construction.

\begin{figure}[htb!]
\begin{minipage}{8in}
\includegraphics[height=3.in]{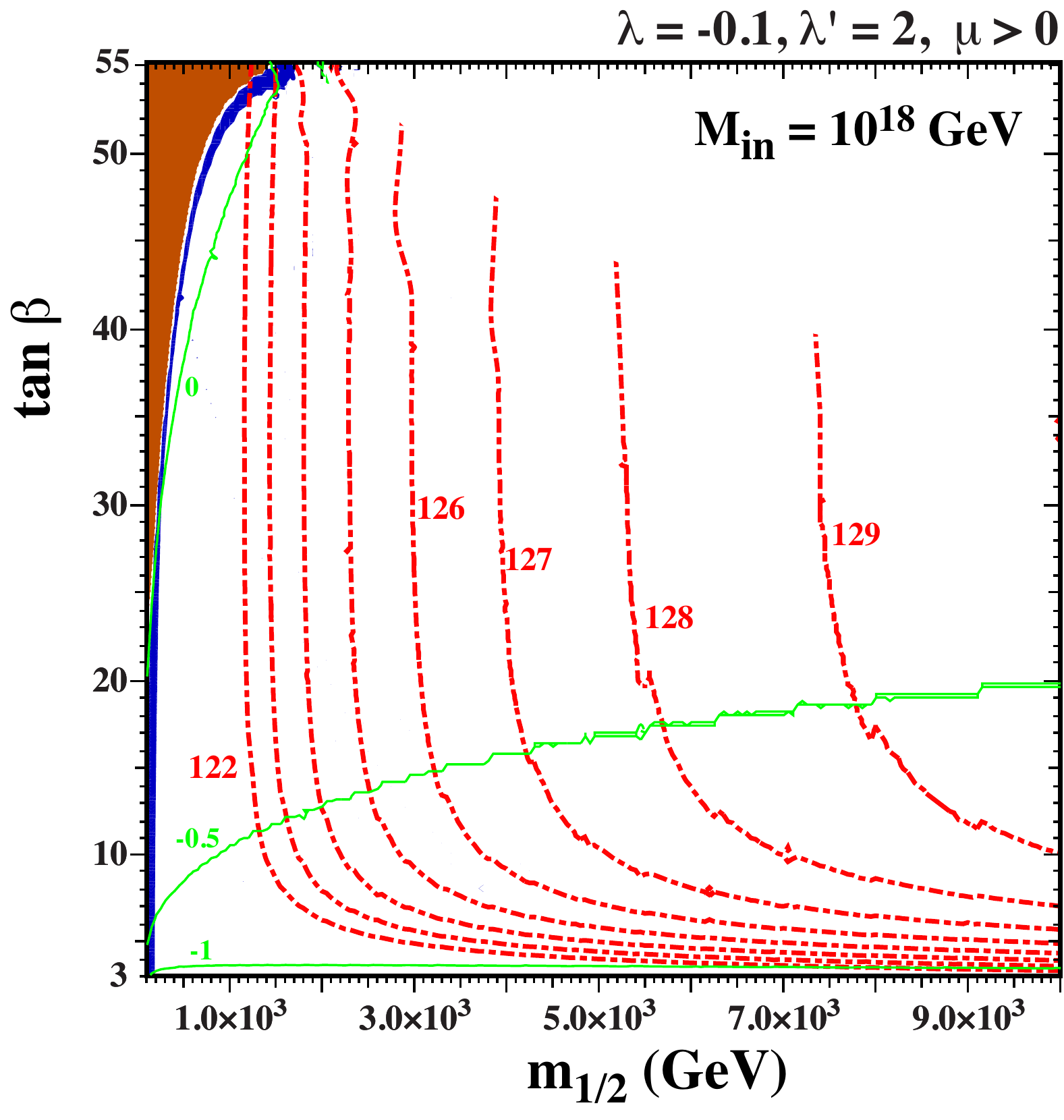}
\includegraphics[height=3.in]{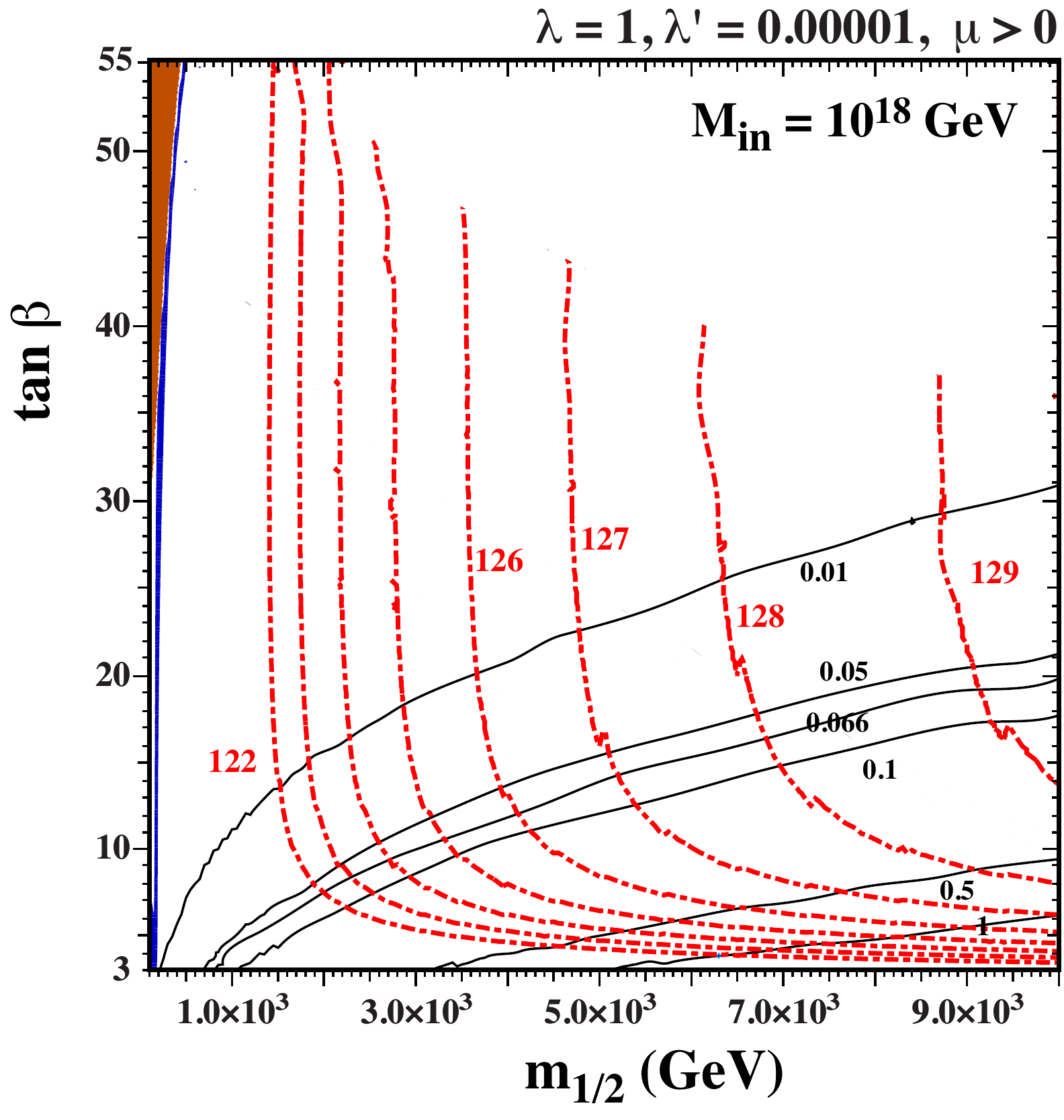}
\hfill
\end{minipage}
\caption{
{\it
Sample no-scale super-GUT $(m_{1/2}, \tan \beta)$ planes for $M_{\rm in
 } = 10^{18}$ GeV. In the left panel $\lambda = -0.1$ and
 $\lambda^\prime = 2$, whereas in the right panel $\lambda = 1$ and
 $\lambda^\prime = 10^{-5}$. The brown shaded region has a stau LSP. The regions compatible with the relic
 density determined by Planck and other experiments are shaded dark
 blue, and the red dot-dashed curves are contours of constant Higgs mass
 as calculated using {\tt FeynHiggs}, which does not give stable results
 in the upper right portions of the panels. In the left panel, the green
 curves are contours of $c_\Sigma (m_{3/2}/m_{1/2})^2$ and the proton
 lifetime is too short throughout. In the right panel, the solid black
 contours show the proton lifetime in units of $10^{35}$ yrs, which is
 acceptably long below the contour labelled 0.066. However, the
 relic density is too large throughout this region.}}
\label{fig:superGM}
\end{figure}

We show in Fig.~\ref{fig:superGM} two examples of $(m_{1/2}, \tan
\beta)$ planes for fixed $M_{\rm in} = 10^{18}$~GeV. In the left panel,
we have chosen $\lambda = -0.1$ and $\lambda^\prime = 2$. In the dark
blue shaded strip, the neutralino LSP relic density agrees with the
value determined by Planck and other experiments. To its left, in the
brown shaded region the stau is either the LSP or tachyonic. The red
dot-dashed contours show the value of the Higgs mass as computed using
the {\tt FeynHiggs} code~\cite{FH}~\footnote{Note that here we use {\tt
FeynHiggs} version 2.11.3, which gives a slightly lower value of $m_h$
than the version used in \cite{emo2}. In addition, since {\tt FeynHiggs}
does not produce stable results in the upper right portion of the plane,
the Higgs contours terminate in this region.}. As one can see, there is a region
at large $\tan \beta \in 52$--55 for $m_{1/2} \in 1$--1.5~TeV that corresponds
to the preferred region found in \cite{emo2}~\footnote{The slight
differences between these and past results arise mostly because here we
do not force the strong gauge coupling to be equal to the electroweak
couplings at the GUT scale.}. In this region the Higgs mass $\in
122$--124~GeV, which is acceptable given the uncertainty in the mass
calculated using {\tt FeynHiggs}. By including a GM term, we are able to
probe lower values of $\tan \beta$ for the same set of input
parameters. Unfortunately, the proton lifetime is much too small over
the entire left panel, with a value of only $10^{25}$~yrs in the upper
left corner. We also show (in green) the contours of the GM term. In
this case, since $|\lambda|\lesssim \lambda^\prime$, we assume $c_H = 0$ and
show the contours of $c_\Sigma (m_{3/2}/m_{1/2})^2$~\footnote{We make no
specific assumption about the magnitude of $m_{3/2}$, except that it is
large enough for the LSP to be the lightest neutralino, rather than the
gravitino.}. As one can see, the contour for $c_\Sigma = 0$
runs through the region of good relic density and Higgs mass found in
\cite{emo2}.

In the right panel of Fig.~\ref{fig:superGM} we show a similar plane but with
different choices of $(\lambda, \lambda^\prime) = (1, 10^{-5})$, which
are more typical of the values required in \cite{eemno}. In this case,
with $\lambda \gg \lambda^\prime$, the value of $c_\Sigma
(m_{3/2}/m_{1/2})^2$ is very near $-0.25$ all across the plane. As long as $c_H$
is relatively small, one can see from Eq.~(\ref{shift}) that the value of $c_H$ has little
effect on our estimate of
$c_\Sigma (m_{3/2}/m_{1/2})^2$ which are quoted assuming $c_H = 0$.
The large ratio of
$\lambda/\lambda^\prime$ is beneficial for increasing the proton
lifetime, and contours showing the lifetime are seen as solid black
curves in the lower right portion of the panel, labelled in units of
$10^{35}$ yrs~\footnote{Details of the calculation of proton decay rates
can be found in Refs.~\cite{eelnos, eemno, evno, Hisano:2013exa}. Here,
we have take the phases in the GUT Yukawa couplings \cite{Ellis:1979hy}
such that the proton decay rate is minimized \cite{eemno}, which gives a
conservative constraint on the model parameter space. }; as the current
experimental limit is $\tau (p\to K^+ \overline{\nu}) > 6.6\times
10^{33}$~yrs \cite{Takhistov:2016eqm}, the region with acceptable proton
stability lies below the contour labelled 0.066. Whilst it is
encouraging that some region of parameter space exists with a
sufficiently long proton lifetime and acceptable Higgs mass,
the relic density is far too large in this region: $\Omega h^2 \sim
\mathcal{O}(100)$. Further exploration in the ($M_{\rm in}, \lambda,
\lambda^\prime$) parameter space does not yield better results. The
Higgs mass can be made compatible with either the relic density or the
proton lifetime, but not both.

The left panel of Fig.~\ref{fig:superGM} shows that, at fixed $m_{1/2}$,
the value of $m_h$ decreases rapidly when $\tan \beta \lesssim 10$. On
the other hand, the right panel of Fig.~\ref{fig:superGM} shows that the
proton lifetime is unacceptably short for $\tan \beta \gtrsim 10$. As we
discuss below with several examples, these two problems can be avoided
simultaneously when $\tan\beta = 7$, for suitable choices of the other
super-GUT model parameters $M_{in}, \lambda$ and $\lambda^\prime$. We do
not discuss in the following possible variations in the value of $\tan
\beta$, but have checked that values differing from 7 by factors
$\gtrsim 2$ are typically excluded by either $m_h$ or the proton
lifetime.

\subsection{Twisted $H$ and $\overline{H}$ Higgs Fields}
\label{sec:twisted}

In this subsection we consider departures from the minimal model discussed above
that allow for more successful phenomenology. We start by considering
the consequences of a twisted Higgs sector. As discussed above, $\tan
\beta$ must be relatively low to obtain sufficiently long proton
lifetimes. However, in order to obtain a sufficiently large Higgs mass,
$\tan \beta$ should not be too low.  Choosing $\tan \beta = 7$ with
$\lambda^\prime = 10^{-5}$ optimizes both $m_h$ and $\tau_p$, so we fix
those values for now.  In the following, we take $\lambda = 0.6$ and 1.

The superpotential (\ref{w_phi}) does not cover the case where twisted
fields couple to untwisted fields. If the Higgs 5-plets are twisted,
then $W_3$ contains Yukawa couplings between the twisted Higgses and
untwisted matter fields. In addition, if $\Sigma$ remains untwisted,
then $W_3$ also contains a term coupling one untwisted field ($\Sigma$)
and the twisted Higgs fields. We define weights for each of the terms in $W_3$:
$\alpha_t$, $\alpha_b$, $\alpha_\lambda$, and $\alpha_\lambda'$
corresponding to the top and bottom Yukawa couplings, the coupling of
the Higgs adjoint to the 5-plets, and the adjoint trilinear,
respectively. Similarly, we define separate weights $\beta_H$ and
$\beta_\Sigma$ for the two bilinears in $W_2$. When both  $H$ and
$\overline{H}$ are twisted, $A$ and $B$ terms are given at the input
renormalization scale by
\beq
 A_{t,b} = (1-\alpha_{t,b})m_{3/2}~, \qquad A_\lambda = (2-\alpha_\lambda) m_{3/2}~, \qquad
 A_{\lambda'} = -\alpha_{\lambda'} m_{3/2}\,,
\eeq
and
\beq
B_H = (2-\beta_H) m_{3/2}~, \qquad  B_\Sigma = -\beta_\Sigma m_{3/2} \, .
\eeq
The Higgs soft squared masses are given by $m_{3/2}^2$ in addition to the usual
supersymmetric contribution from $\mu$ (properly shifted by the GM term).

\begin{figure}[htb!]
\begin{minipage}{8in}
\includegraphics[height=3.in]{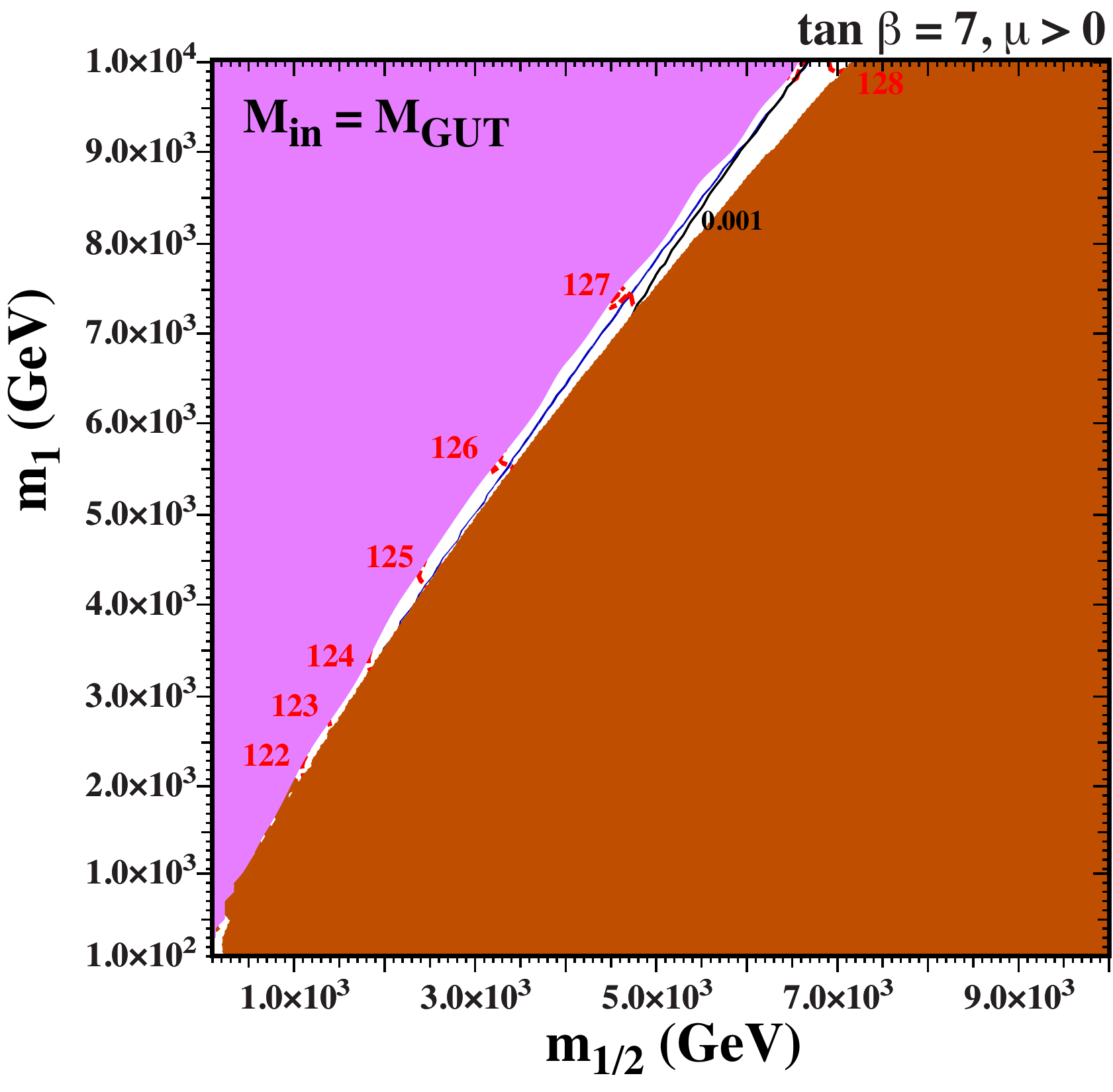}
\includegraphics[height=3.in]{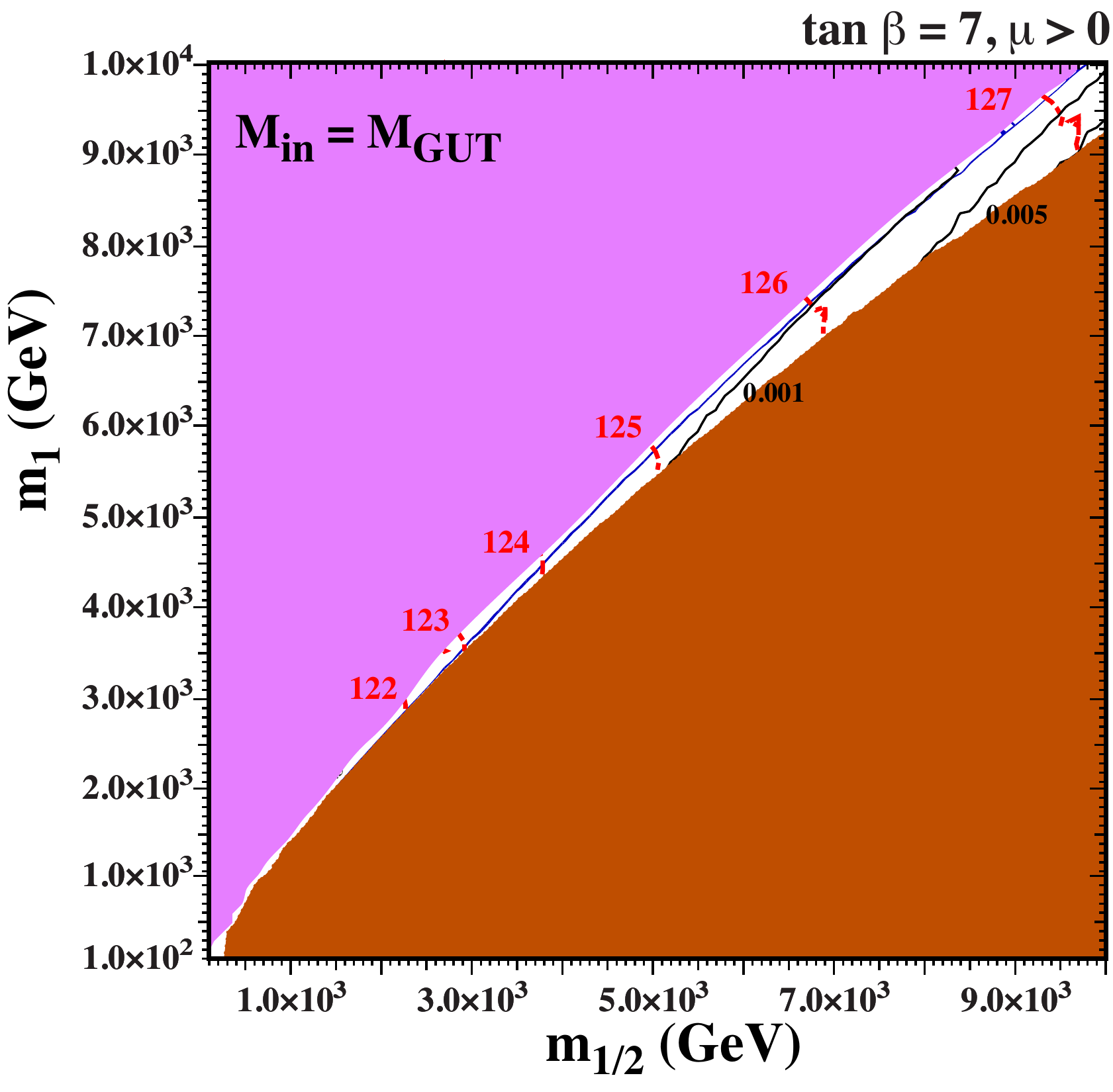}
\hfill
\end{minipage}
\caption{
{\it
Examples of  $(m_{1/2}, m_1)$ planes for $M_{\rm in } = M_{\rm GUT}$ and
$\tan \beta = 7$ when both Higgs 5-plets are twisted. In the left panel,
all the modular weights $\alpha_i = \beta_i = 0$, corresponding to $A_{t,b} = m_1$,
$A_\lambda = 2 m_{3/2}$, $A_{\lambda'} = 0$, $B_H = 2 m_1$, and $B_\Sigma = 0$.
In the right panel, the modular weights are chosen to be $\alpha_{t,b} = 1$,
$\alpha_\lambda = 2$, $\alpha_{\lambda'} = 0$, $\beta_H = 2$
and $\beta_\Sigma = 0$, corresponding to $A_0 = B_0 = 0$. The shadings
and contour colours are the same as in Fig.~\protect\ref{fig:superGM}.
The pink shaded region corresponds to parameter
choices where the electroweak vacuum conditions cannot be satisfied and radiative electroweak
symmetry breaking is not possible.
}}
\label{fig:twist2}
\end{figure}

We consider first the case where both Higgs 5-plets are twisted, and therefore
receive equal soft supersymmetry breaking masses, $m_1 = m_2 =
m_{3/2}$. We start by taking all of the modular weights $\alpha = \beta
= 0$ as before. Now, however, there are non-zero $A$ and $B$
terms at the input scale. We assume  $A_{t,b} = m_{3/2}$, $A_\lambda = 2
m_{3/2}$, $A_{\lambda'} = 0$, $B_H = 2 m_{3/2}$ and $B_\Sigma = 0$ at
the input renormalization scale, $M_{in}$. The ($m_{1/2}, m_1$) plane
for this case with $M_{\rm in} = M_{\rm GUT}$ is shown in the left panel
of Fig.~\ref{fig:twist2}. This is the limiting case in which the
super-GUT scenario reduces to an NUHM1 plane
\cite{nuhm1,eosk,elos,eelnos} with $m_0 = 0$ and $A_0 = m_1$. Note that
the values of $\lambda$ and $\lambda^\prime$ are irrelevant when taking $M_{\rm in} =
M_{\rm GUT}$ as there is no running above the GUT scale in this case.
 There is
narrow band where the LSP is the lightest neutralino and the electroweak
symmetry breaking conditions can be satisfied, through which runs a blue
relic density strip. At low values of $m_{1/2}$, the relic density is
determined by stau coannihilation~\cite{stau}, and the blue relic
density strip lies close to the boundary of the stau LSP region (shaded
red). At higher $m_{1/2}$, the strip moves closer to the region with no
electroweak symmetry breaking (shaded pink) and becomes a focus-point
strip \cite{fp}. The Higgs mass (shown by the red dot-dashed contours
between the two excluded regions) has acceptable values along much of
the relic density strip. On the other hand, the proton lifetime is too
short as the entire strip shown lies at or below the contour
corresponding to $\tau_p = 0.001 \times 10^{35}$~yrs (which appears as
the black curve that enters the allowed region at about 5~TeV at an
angle to the relic density strip). The right panel of
Fig.~\ref{fig:twist2} shows the corresponding plane with the following
choices of modular weights: $\alpha_{t,b} = 1$,
$\alpha_\lambda = 2$, $\alpha_{\lambda'} = 0$, $\beta_H = 2$
and $\beta_\Sigma = 0$, which correspond to $A_0 = B_0 = 0$.
This exhibits many features similar to the left panel. In particular,
the relic density and proton lifetime constraints are incompatible,
motivating our exploration of super-GUT scenarios.

\begin{figure}[htb!]
\begin{minipage}{8in}
\includegraphics[height=3.in]{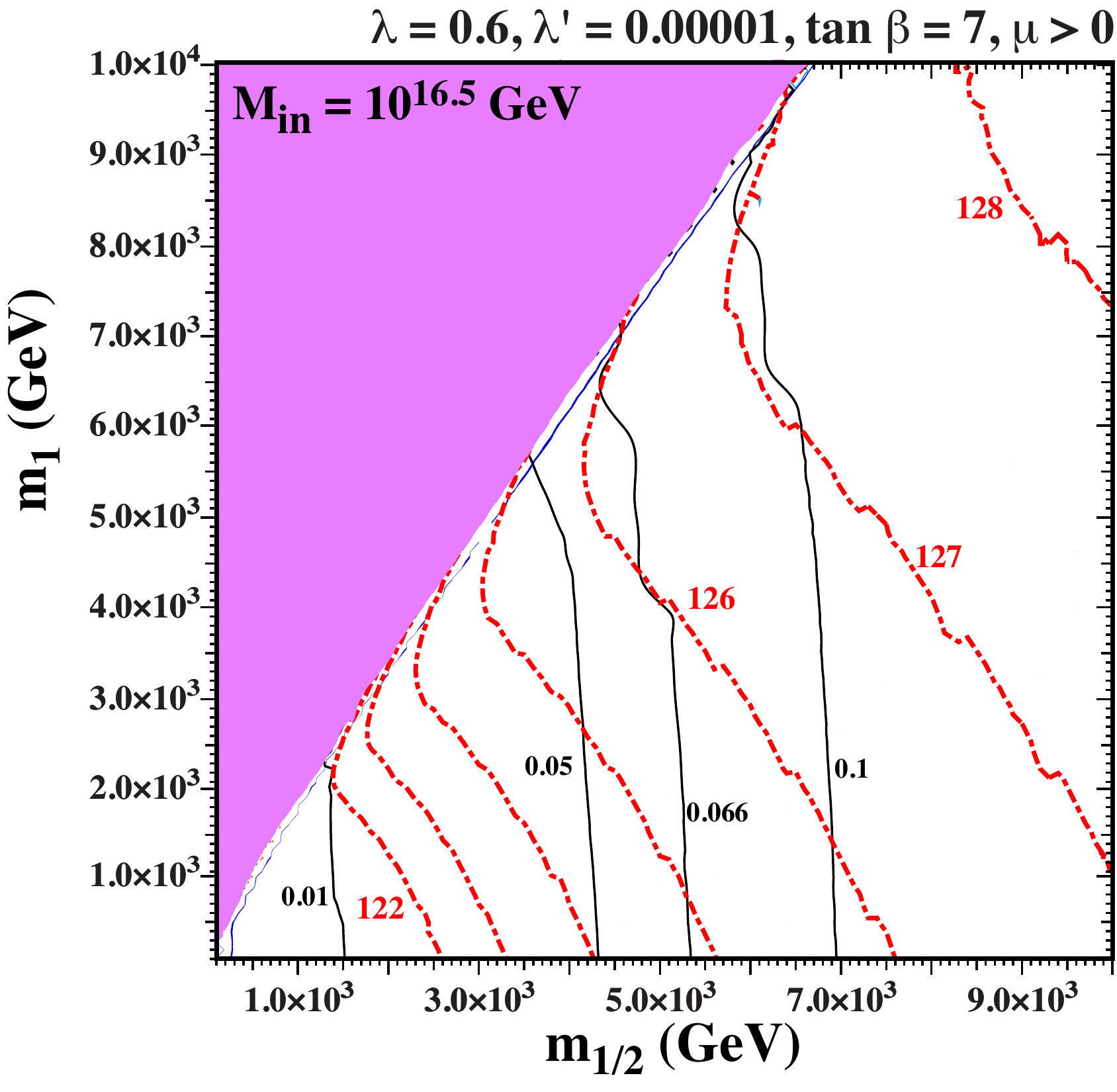}
\includegraphics[height=3.in]{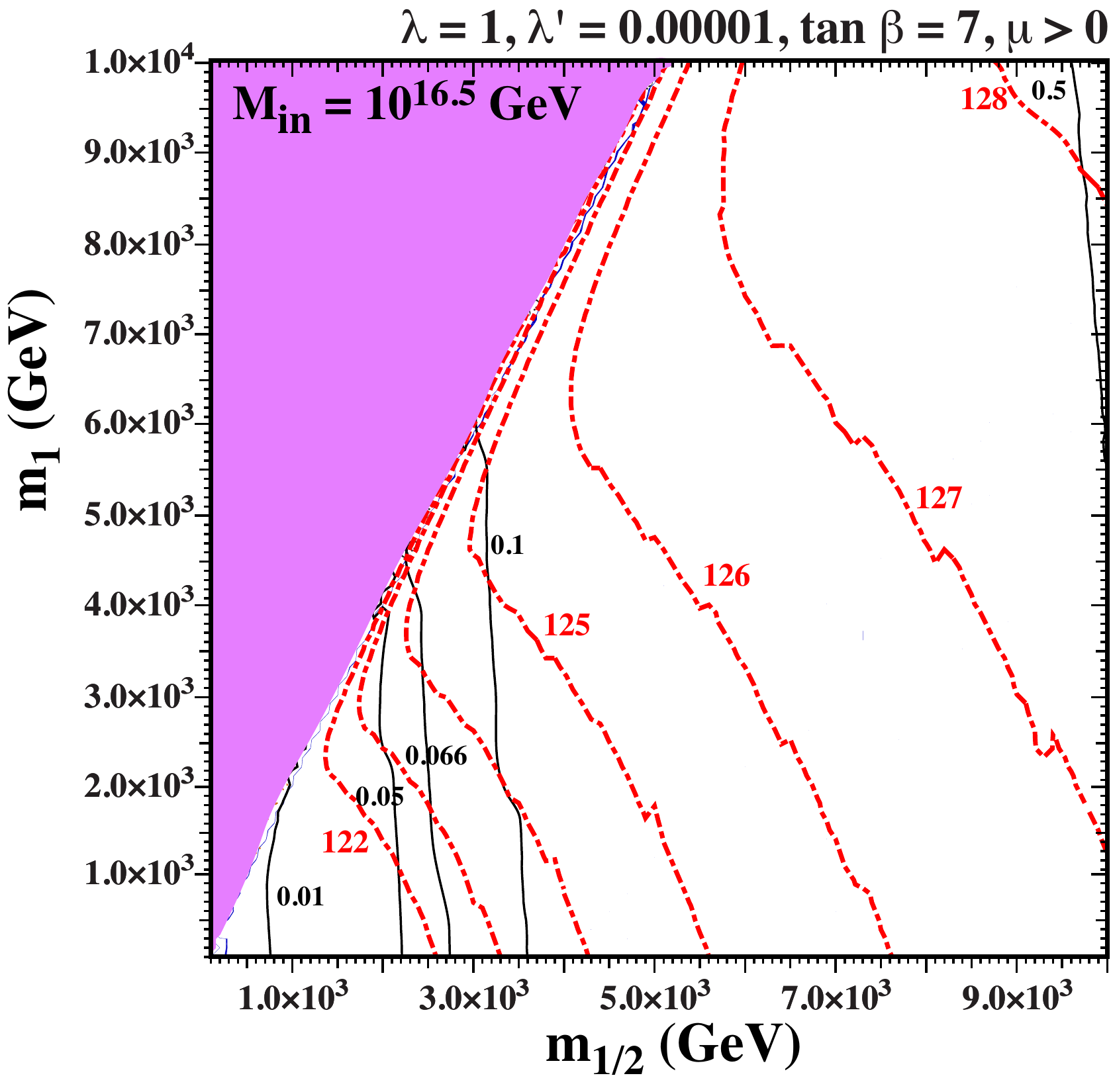}
\hfill
\end{minipage}
\caption{
{\it
Examples of  $(m_{1/2}, m_1)$ planes for $M_{\rm in } = 10^{16.5}$~GeV when both Higgs 5-plets
are twisted.
All the modular weights $\alpha_i = \beta_i = 0$, corresponding to $A_{t,b} = m_1$,
$A_\lambda = 2 m_{3/2}$, $A_{\lambda'} = 0$, $B_H = 2 m_1$, and $B_\Sigma = 0$.
In both panels $\tan \beta = 7$ and $\lambda^\prime = 10^{-5}$ with $\lambda = 0.6$ (left)
and $\lambda = 1$ (right). The shadings
and contour colours are the same as in Fig.~\protect\ref{fig:superGM}. The pink shaded region corresponds to parameter
choices where the electroweak vacuum conditions cannot be satisfied, and radiative electroweak
symmetry breaking is not possible.
}}
\label{fig:twist21}
\end{figure}

In Fig.~\ref{fig:twist21}, the model with all weights set to zero is
assumed again, but now with $M_{\rm in} = 10^{16.5}$~GeV. The most
dramatic difference between this model and the previous GUT model shown
in the left panel of Fig.~\ref{fig:twist2} is the disappearance of the
stau LSP region as $M_{\rm in}$ is increased above the GUT scale, an
effect that was discussed in \cite{emo,super-GUT}. In the super-GUT
case even a small amount of running with $m_0 = 0$ between $M_{\rm GUT}$
and $M_{\rm in}$ is sufficient to restore a neutralino LSP. In the left
panel of this figure, we have taken the Higgs coupling, $\lambda = 0.6$,
whereas in the right panel $\lambda = 1$, fixing $\lambda^\prime =
10^{-5}$ in both panels.  In this case, the relic density strip (which
is little changed from the GUT model) lies close to the boundary where
electroweak symmetry breaking  is not possible (shaded pink),  and is
similar to the focus-point region of the CMSSM~\cite{fp}.

Another very obvious difference between the left panel of
Fig.~\ref{fig:twist2} and Fig.~\ref{fig:twist21} is the value of the
proton lifetime. With $M_{\rm in} = M_{\rm GUT}$, the entire strip shown
has a lifetime $\tau_p < 10^{33}$~yrs, as it lies to the left of the
contour labeled 0.01. However, the proton lifetime is significantly
longer in both panels of Fig.~\ref{fig:twist21}, and there are
acceptable parts of the relic density strip where $\tau_p > 0.066 \times
10^{35}$~yrs. Comparing the two panels allows one to see the effect of
increasing $\lambda$ on the proton lifetime. For $\lambda = 0.6$, the
lifetime is  sufficiently long for $m_{1/2} \gtrsim 5$~TeV, whereas for
$\lambda = 1$ this is relaxed to $m_{1/2} \gtrsim 2.5$~TeV. Increasing
$\lambda$ much further is not possible due to its effect on the Yukawa
couplings, as discussed in \cite{eemno}. In both cases, the Higgs masses
are reasonably consistent with 125~GeV, though due to the increased
``bending'' of the contours, the Higgs mass along the relic density
strip is slightly lower for the larger value of $\lambda$~\footnote{At
higher $M_{in}$, the bending of Higgs mass contours seen in
Fig.~\ref{fig:twist21} as they approach the region with no radiative
electroweak symmetry breaking (shaded pink) becomes more severe, and the
Higgs mass becomes too low all along the relic density strip.}.
The GM couplings are also acceptably small: in the GUT case shown in the
left panel of Fig. \ref{fig:twist2} they are $\ll 1$ across the plane,
whereas in Fig.~\ref{fig:twist21} $c_\Sigma (m_{3/2}/m_{1/2})^2$ is of
order 0.05 all along the relic density strip.

Since the strips with acceptable relic density in these models
resemble the familiar focus-point region~\cite{fp}, one can expect that the spin-independent elastic scattering
cross section on protons, $\sigma^{\rm SI}$, may be relatively large.  Concentrating on the
right panel of Fig.~\ref{fig:twist21}, we have computed
$\sigma^{\rm SI}$ at two points:
($m_{1/2}, m_1$) = (3100,6000) GeV and (4100, 8000) GeV.
The resulting cross sections are $\sigma^{\rm SI} = (1.24 \pm 0.77)
\times 10^{-8}$~pb and $(1.90 \pm 1.19) \times 10^{-9}$~pb with $m_\chi = 930$ GeV and 1400 GeV, respectively,
where we have assumed $\Sigma_{\pi N} = 50 \pm 8$~MeV \cite{eosv} and
$\sigma_0 = 36\pm 7$~MeV \cite{Borasoy:1996bx}.
The central value for the former point is slightly above the recent LUX \cite{lux} and PandaX \cite{pandax}
bounds, but remains acceptable when uncertainties in the computed cross sections are taken into
account.  Furthermore, using nucleon matrix elements computed with
lattice simulations as in~\cite{EMT} would reduce the predicted cross
section by more than a factor 2 due to the smallness of strange-quark
content in a nucleon.
However, in both the cases studied one may anticipate a positive signal in upcoming direct detection experiments
such as LUX-Zeplin and XENON1T/nT~\cite{futureDD}.

\begin{figure}[htb!]
\vspace*{0.5cm}
\begin{minipage}{8in}
\includegraphics[height=3.in]{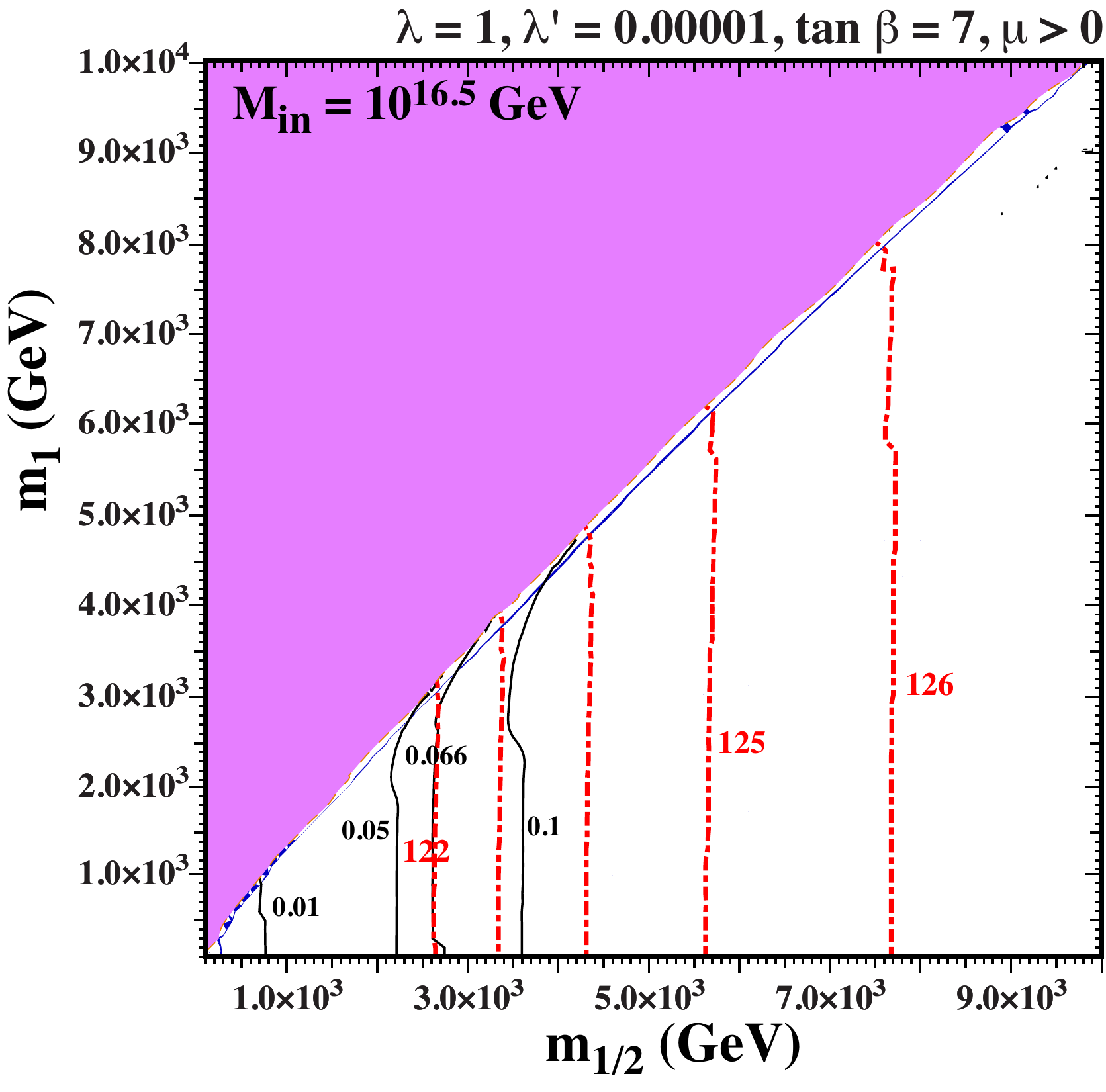}
\includegraphics[height=3.in]{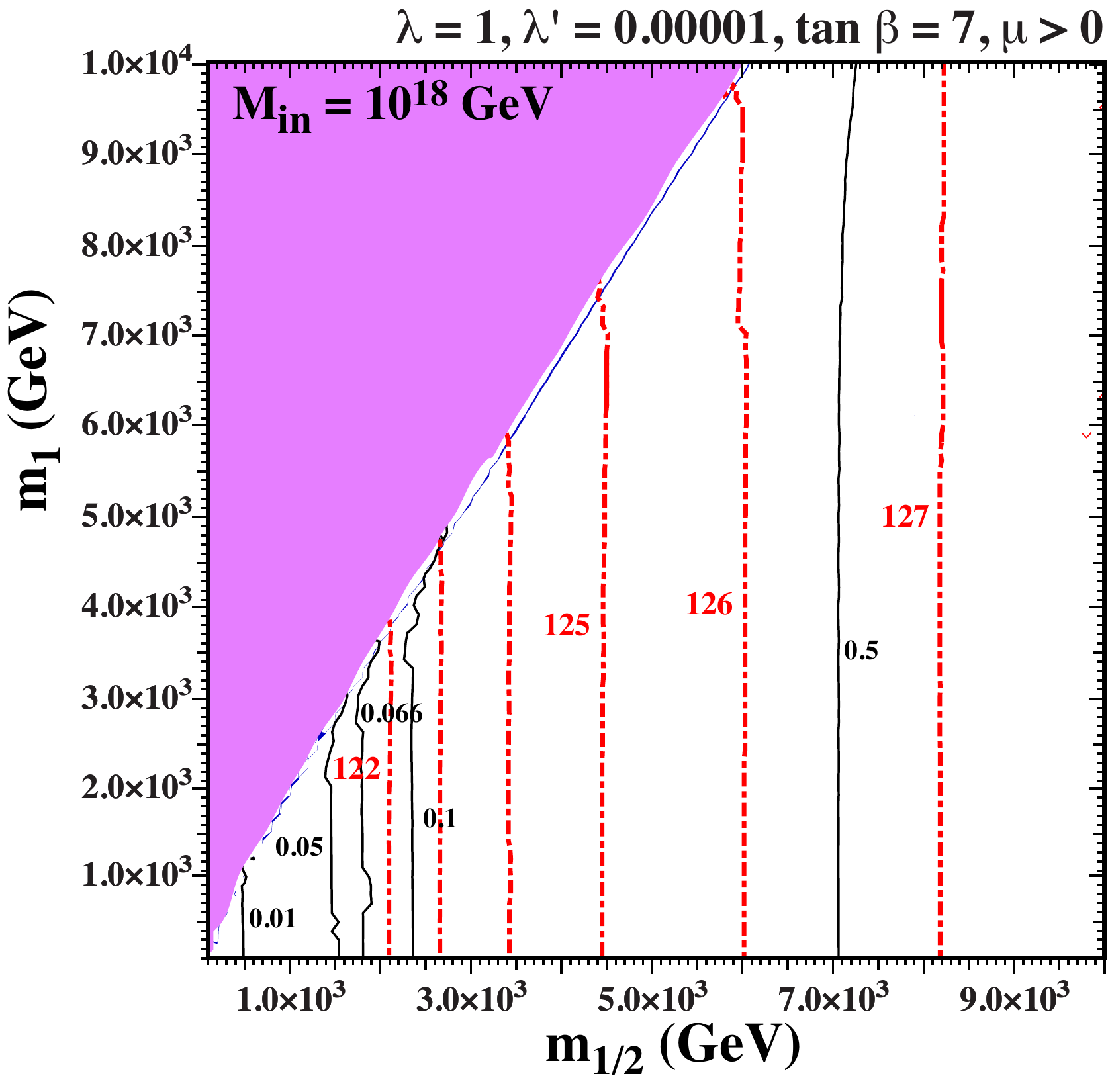}
\hfill
\end{minipage}
\caption{
{\it
Examples of  $(m_{1/2}, m_1)$ planes for $M_{\rm in } = 10^{16.5}$
 (left) and $10^{18}$~GeV (right) when both Higgs 5-plets are twisted. In both cases the modular weights are
 $\alpha_{t,b} = 1$, $\alpha_\lambda = 2$, $\alpha_{\lambda'} = 0$,
 $\beta_H = 2$ and $\beta_\Sigma = 0$, corresponding to $A_0 = B_0 = 0$,
 and we assume $\tan \beta = 7$, $\lambda = 1$ and $\lambda^\prime =
 10^{-5}$. The shadings and contour colours are the same as in
 Fig.~\protect\ref{fig:superGM}.
}}
\label{fig:twist22}
\end{figure}

We consider next the case with the modular weights $\alpha_{t,b} = 1$,
$\alpha_\lambda = 2$, $\alpha_{\lambda'} = 0$, $\beta_H = 2$ and
$\beta_\Sigma = 0$, so that $A_0 = B_0 = 0$ for all $A$ and $B$
terms. The right panel of Fig.~\ref{fig:twist2} shows the $(m_{1/2},
m_1)$ plane for $M_{\rm in} = M_{\rm GUT}$, which is similar to that
shown in the left panel when $A$ and $B$ terms are non-zero. The $A$ and
$B$ terms are seen to affect somewhat the dependence on $m_1$ of the
Higgs mass and the position of the relic density strip. The same case
with $A_0 = B_0 = 0$ but $M_{\rm in} = 10^{16.5}$~GeV is shown in the
left panel of Fig.~\ref{fig:twist22}. Comparing this with the right
panel of Fig.~\ref{fig:twist21}, we see that the proton lifetime shows
little dependence on $A_0$ and is similar in the two cases shown. For
larger $M_{\rm in} = 10^{18}$~GeV with $A_0 = B_0 = 0$, as shown in the
right panel of Fig.~\ref{fig:twist22}, we see that the relic density
strip shifts to larger values of $m_1$ and the proton lifetime is
somewhat longer.  Much of the allowed dark matter strip has an
acceptably long proton lifetime. The effect of adjusting the modular
weights does not have a major effect on the elastic scattering cross section.

We consider next the case where only one of the Higgs 5-plets is twisted,
so that
\beq
 A_\lambda = (1-\alpha_\lambda) m_{3/2} \qquad
 A_{\lambda'} = -\alpha_{\lambda'} m_{3/2} \qquad
B_H = (1-\beta_H) m_{3/2} \qquad  B_\Sigma = -\beta_\Sigma m_{3/2} \, .
\eeq
When $\overline{H}$ is twisted,
\beq
A_t = -\alpha_t m_{3/2} \qquad A_b = (1-\alpha_b) m_{3/2} \, ,
\eeq
whereas when $H$ twisted,
\beq
A_b = -\alpha_b m_{3/2} \qquad A_t = (1-\alpha_t) m_{3/2} \, .
\eeq
Thus, in either case we have non-universal $A$-terms related via the
Yukawa couplings.

\begin{figure}[htb!]
\begin{minipage}{8in}
\includegraphics[height=3.in]{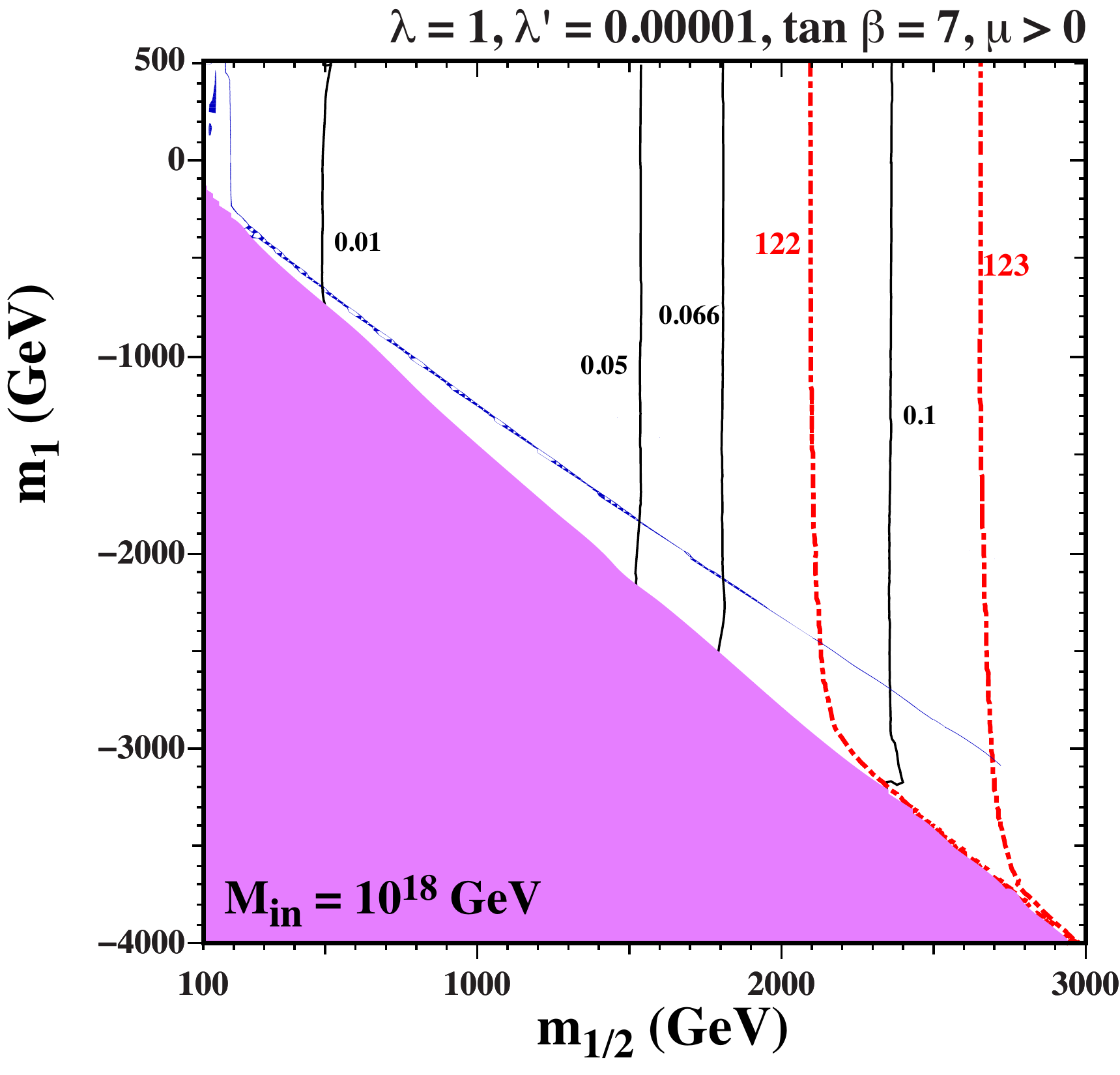}
\includegraphics[height=3.in]{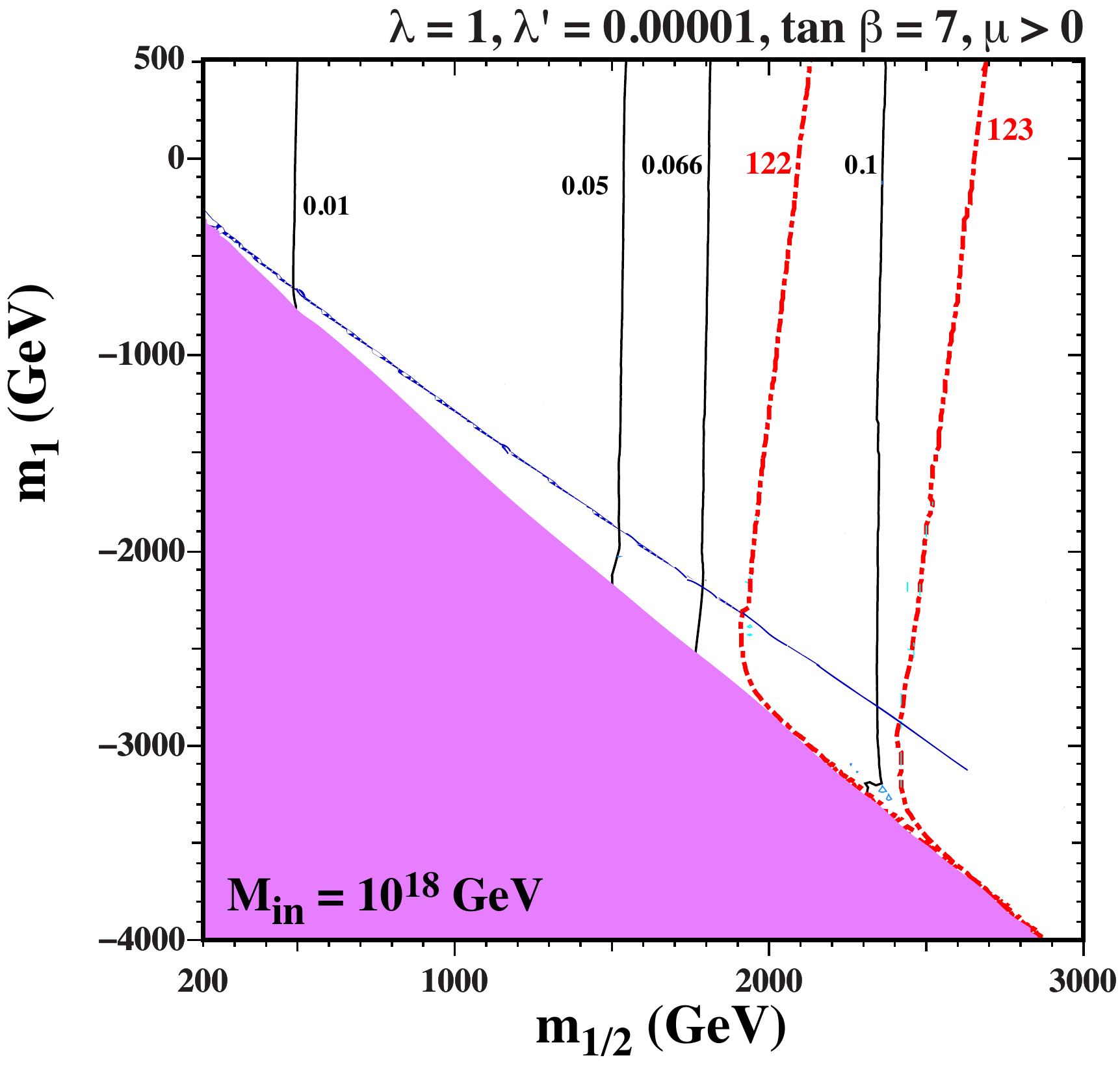}
\hfill
\end{minipage}
\caption{
{\it
Examples of  $(m_{1/2}, m_1)$ planes for $M_{\rm in } = 10^{18}$~GeV
when only $\overline{H}$ is twisted.
In the left panel, the modular weights are $\alpha_{t} = 0$, $\alpha_b =
 1$, $\alpha_\lambda = 1$, $\alpha_{\lambda'} = 0$, $\beta_H = 1$ and
 $\beta_\Sigma = 0$, so that all trilinear and bilinear terms vanish. In
 the right panel, all the weights vanish, so that $A_t = 0$, $A_b =
 m_1$, $A_\lambda = m_1$, $A_{\lambda'} = 0$, $B_H = m_1$, and $B_\Sigma
 = 0$. In both panels $\tan \beta = 7$,  $\lambda = 1$ and
 $\lambda^\prime = 10^{-5}$. The shadings and contour colours are the
 same as in Fig.~\protect\ref{fig:superGM}.
}}
\label{fig:twist11}
\end{figure}

We consider first the case with twisted $\overline{H}$. Examples of
$(m_{1/2}, m_1)$ planes for $M_{\rm in } = 10^{18}$~GeV are shown in
Fig.~\ref{fig:twist11}. In both panels, we have taken $\tan \beta = 7$,
$\lambda = 1$, and $\lambda^\prime = 10^{-5}$. Since $H$ remains
untwisted, we have $m_0 = m_2 = 0$ and, since the two Higgs soft masses
are unequal, this is an example of a super-GUT NUHM2 model \cite{nuhm2, eosk,
elos, eelnos}~\footnote{The quoted sign of $m_1$ actually represents the sign of
$m_1^2$ at the input scale.}. The region where one obtains an acceptable
relic density could be expected from the upper left panel of Fig.~14 in
\cite{eosk}, which shows an example of an $(m_1,m_2)$ plane for relatively low
$m_{1/2}$, $m_0$ and $\tan \beta$. For $m_2 = 0$, we expect that there
should be a funnel strip \cite{funnel} where $s$-channel annihilation of
the LSP through the heavy Higgs scalar and pseudoscalar dominates the
total cross section and $m_\chi \approx m_A/2$. This generally occurs
when $m_1^2 < 0$ at the input scale.

In the left panel of Fig.~\ref{fig:twist11}, we have taken $\alpha_t =
0$, $\alpha_{b} = 1$, $\alpha_\lambda = 1$, $\alpha_{\lambda'} = 0$,
$\beta_H = 1$ and $\beta_\Sigma = 0$, so that all the $A$ and $B$ terms
vanish at the input scale. In the pink shaded region, the electroweak
symmetry breaking (EWSB) conditions cannot be satisfied as $m_A^2 <
0$. Indeed, for $m_1^2 < 0$, we see a blue relic density strip above the
shaded region. Whilst the proton lifetime is sufficiently large for
$m_{1/2} \gtrsim 1.8$~TeV, the strip extends (barely visibly) to $m_h =
123$~GeV (shown by the red dot-dashed contours). In the right panel of
this figure, we have set all weights to zero, and therefore $A_t = 0$,
$A_b = m_1$, $A_\lambda = m_1$, $A_{\lambda'} = 0$, $B_H = m_1$, and
$B_\Sigma = 0$. Qualitatively, the two figures are very similar. The
strip extends to slightly larger $m_h$ but, again, not much past
123~GeV. In both cases, $A_t = 0$ at the input scale and, although $A_b
\ne 0$ in the right panel, the dominant factor contributing to the Higgs
mass is $A_t$. In both panels $c_\Sigma (m_{3/2}/m_{1/2})^2 \approx
-0.25$ in the allowed regions of the parameter space. We see that the
proton lifetime is acceptably long when $m_{1/2} \gtrsim 1.7$~TeV
along the dark matter strip. The elastic cross section near the end point of the
relic density strip where $m_1 \approx -3500$ GeV is quite small:
$\sigma^{\rm SI} \approx 1 \times 10^{-11}$ pb with $m_\chi \approx 800$ GeV,
probably beyond the reach of LUX-Zeplin and XENON1T/nT~\cite{futureDD}, though still
above the neutrino background level.

\begin{figure}[htb!]
\vspace*{0.5cm}
\begin{minipage}{8in}
\includegraphics[height=3.in]{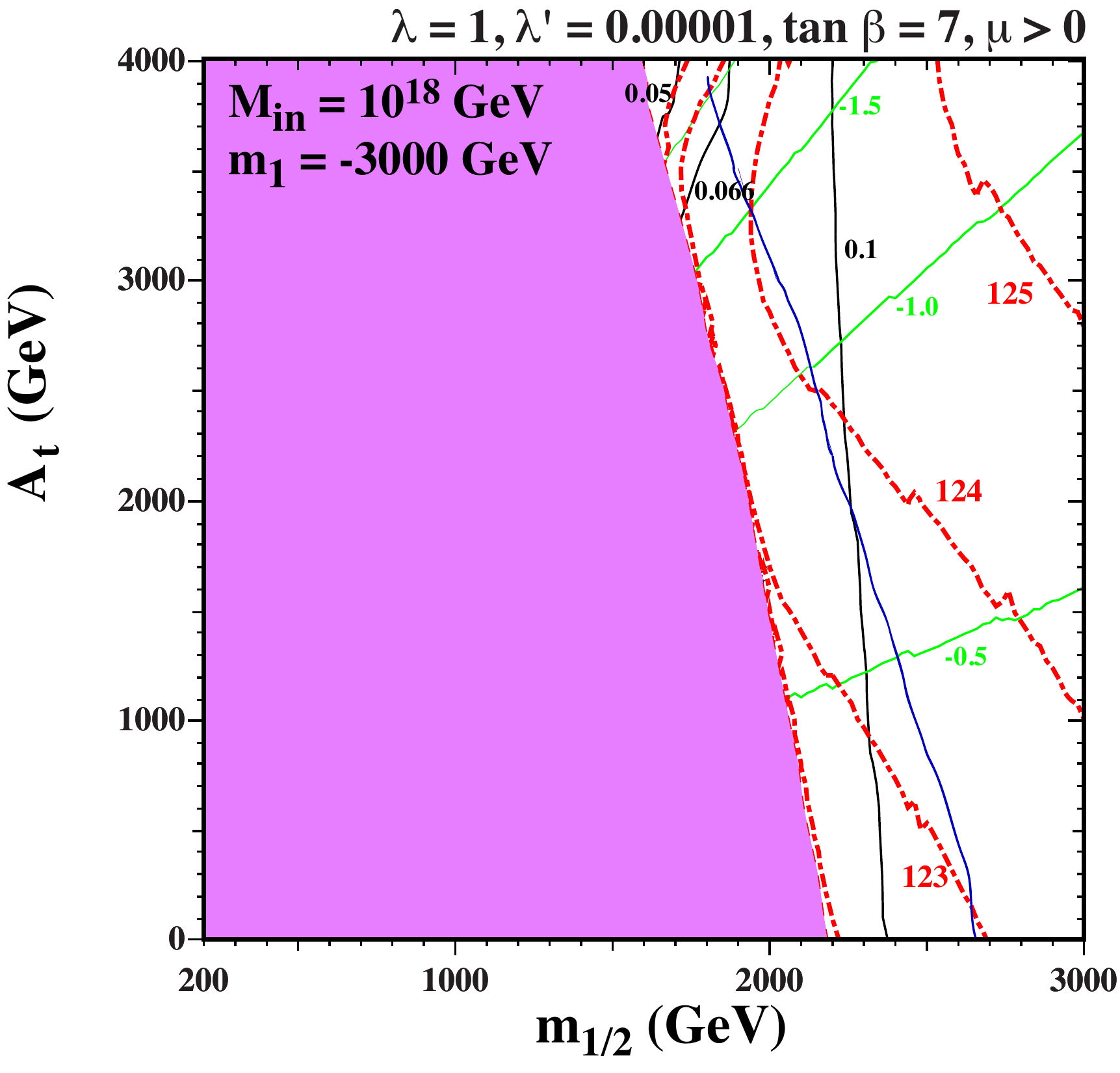}
\includegraphics[height=3.in]{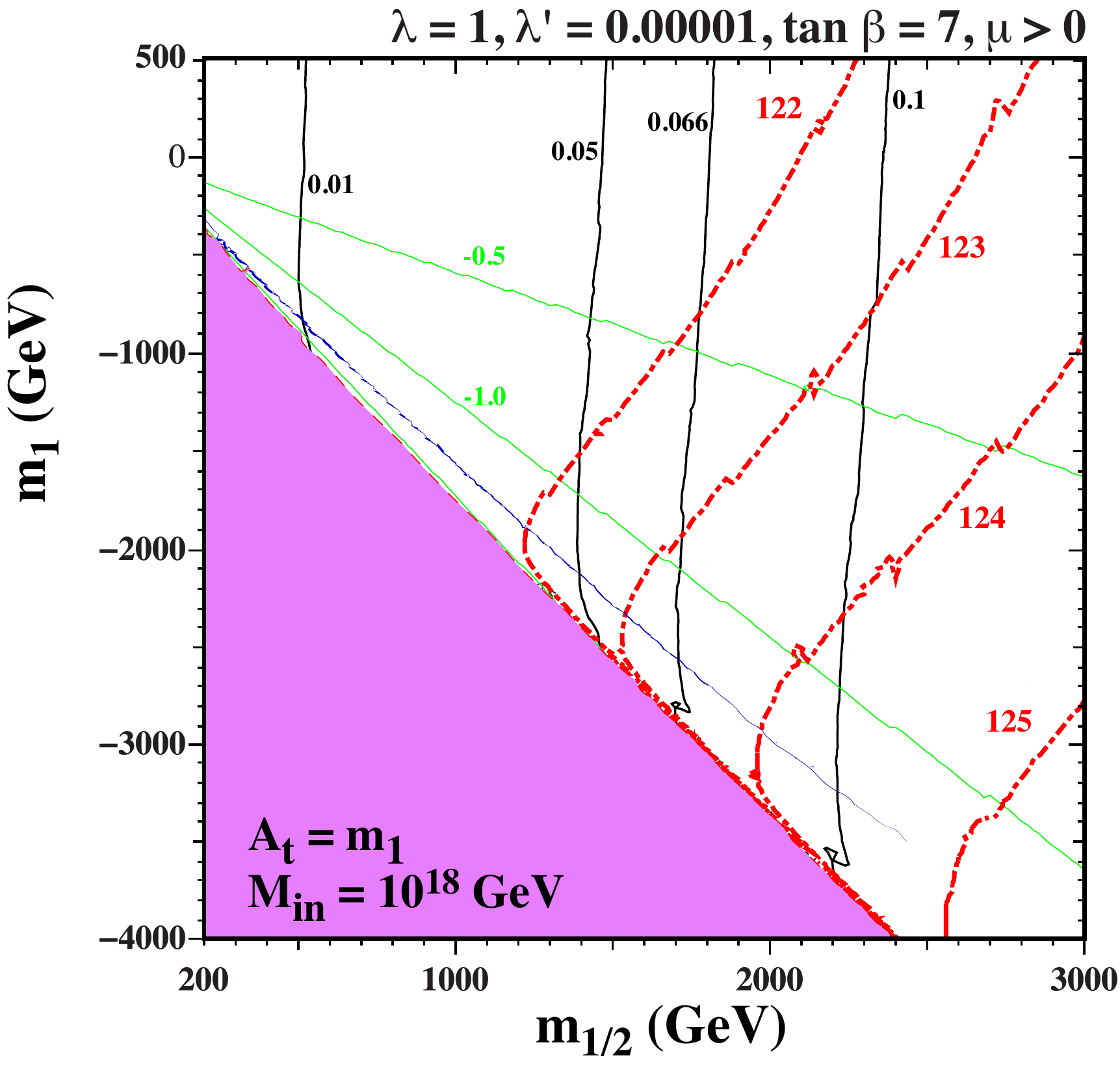}
\hfill
\end{minipage}
\caption{
{\it
Left panel: The $(A_t, m_{1/2})$ plane for $\tan \beta = 7,  \lambda = 1$ and $\lambda^\prime = 10^{-5}$
with $M_{\rm in} = 10^{18}$ GeV, $m_0 = 0$ and $m_1 = -3000$ GeV when only $\overline{H}$ is twisted.
Right panel:
The corresponding $(m_{1/2}, m_1)$ plane.
Here, the modular weights are $\alpha_b = 1$, $\alpha_\lambda = 1$, $\alpha_{\lambda'} = 0$, $\beta_H = 1$
and $\beta_\Sigma = 0$. For the left panel $\alpha_t$ varies and $m_1$ is fixed while for the right panel $A_t = m_1$ ($\alpha_t = -1$) and all other $A_i = B_i = 0$.
The shadings and contour colours are the same as in Fig. \protect\ref{fig:superGM}.
}}
\label{fig:twist12}
\end{figure}

The Higgs mass can be increased slightly by turning on the weight
$\alpha_t$ controlling $A_t$. To determine the optimal value for
$\alpha_t$, for all other $A_i=0$ and $B_i=0$, we scan over $\alpha_t$. In the left panel of Fig.~\ref{fig:twist12} the resulting $(A_t,
m_{1/2})$ plane for fixed $m_1 = -3000$~GeV and $m_0 = m_2 = 0$ is shown. Once
again, the pink shaded region is excluded as $m_A^2 < 0$ and the
constraints for electroweak symmetry breaking cannot be satisfied.
The blue line (enhanced here for visibility) shows the position of the
relic density funnel strip. We see that the largest value of the Higgs
mass obtained is slightly larger than 124~GeV, which is reached when
$A_t/|m_1| \sim 1$. The proton lifetime is acceptably long
for $m_{1/2} \gtrsim 1.8$~TeV along the dark matter strip, and the
GM coupling shown by the green lines is
$\gtrsim -1.5$ in this region. In the right panel, we show the
corresponding $(m_{1/2}, m_1)$ plane with $A_t = m_1$ and again all other $A_i
= B_i = 0$. Here we see that the funnel strip extends to Higgs masses
slightly larger than 124~GeV, where the proton lifetime is about
$10^{34}$~yrs. Points along the dark matter strip with $m_{1/2} \gtrsim 1.7$~TeV
have an acceptably long proton lifetime.  In both cases, displayed, the elastic cross sections
are relatively small. Near the end point of the relic density strip where $m_1 \approx -3000$ GeV,
we find $\sigma^{\rm SI} \approx 2 \times 10^{-11}$ pb with $m_\chi \approx 950$ GeV.
Although this cross section is still above the neutrino background, it may be difficult to
detect in the planned LUX-Zeplin and XENON1T/nT experiments.

\begin{figure}[htb!]
\begin{minipage}{8in}
\includegraphics[height=3.in]{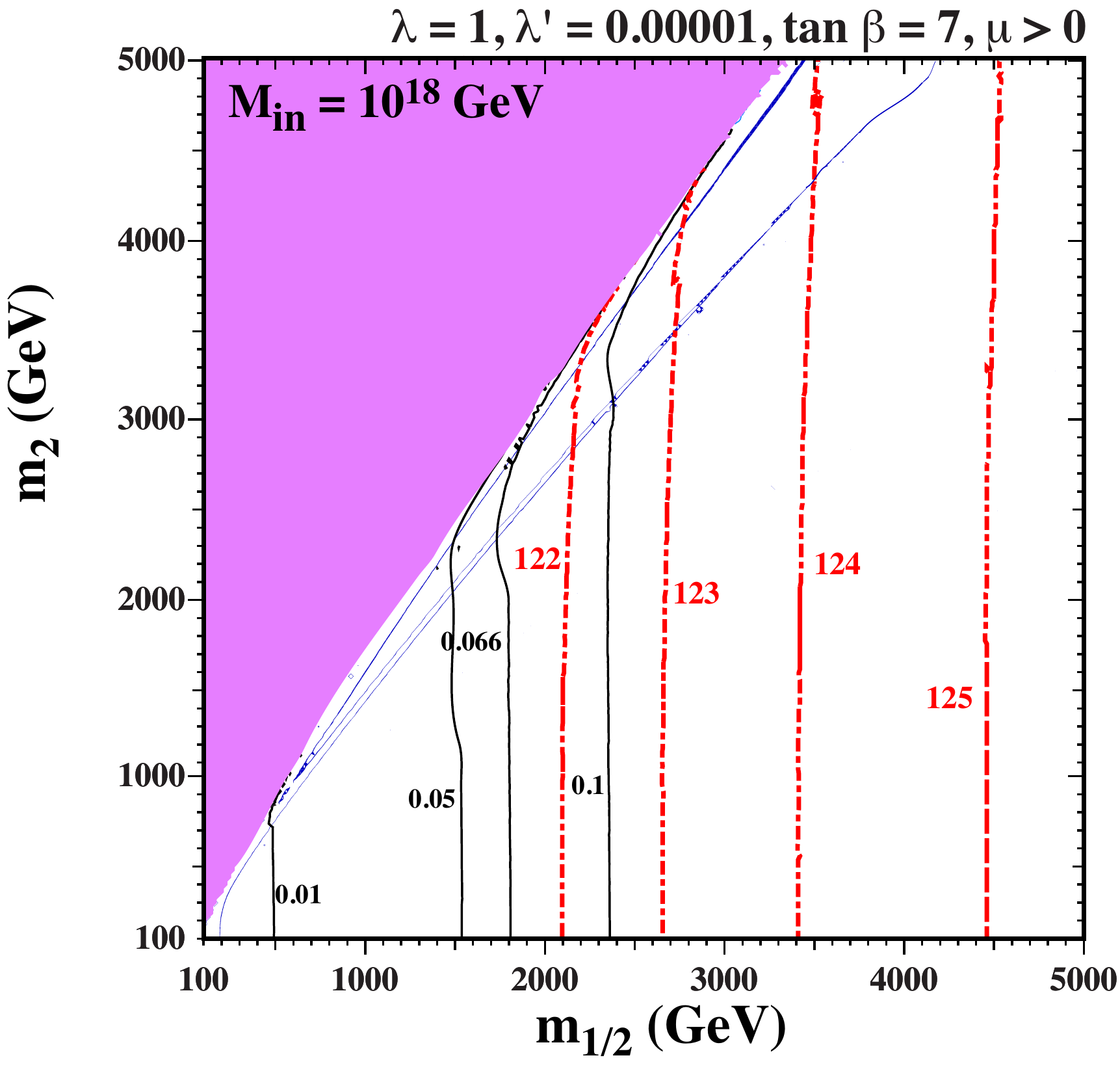}
\includegraphics[height=3.in]{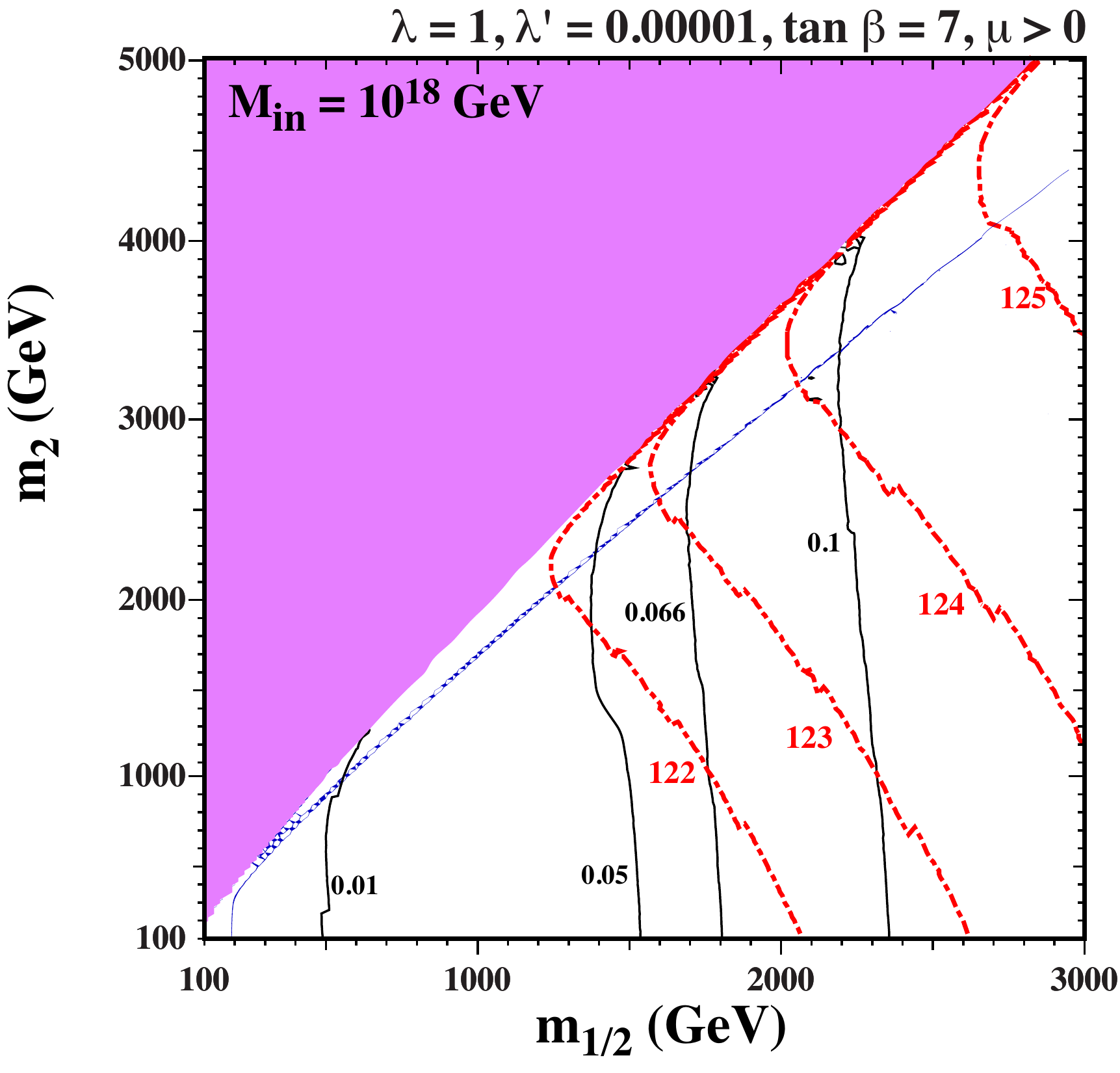}
\hfill
\end{minipage}
\caption{
{\it
Examples of  $(m_{1/2}, m_2)$ planes for $M_{\rm in } = 10^{18}$ when only $H$ is twisted.
In the left panel, all trilinear and bilinear terms are zero.
The modular weights are $\alpha_{t} = 1$, $\alpha_b = 0$, $\alpha_\lambda
 = 1$, $\alpha_{\lambda'} = 0$, $\beta_H = 1$ and $\beta_\Sigma = 0$,
 corresponding to $A_0 = B_0 = 0$. In the right panel, all weights are
 zero, so that $A_t = m_2$, $A_b = 0$, $A_\lambda = m_2$, $A_{\lambda'}
 = 0$,  $B_H = m_2$, and $B_\Sigma = 0$. In both panels $\tan \beta =
 7$, $\lambda = 1$ and $\lambda^\prime = 10^{-5}$. The shadings and
 contour colours are the same as in Fig.~\protect\ref{fig:superGM}.
}}
\label{fig:twist13}
\end{figure}

Finally, we consider the effects of twisting $H$ leaving $\overline{H}$
untwisted. In this case, $m_0 = m_1 = 0$, and previous studies lead us
to expect the relic density strip to lie at positive values of
$m_2^2$. Once again, we have taken $\tan \beta = 7$, $\lambda = 1$, and
$\lambda^\prime = 10^{-5}$. In the left panel of Fig.~\ref{fig:twist13},
we have taken $\alpha_t = 1$, $\alpha_{b} = 0$, $\alpha_\lambda = 1$,
$\alpha_{\lambda'} = 0$, $\beta_H = 1$ and $\beta_\Sigma = 0$, so that
all $A$ and $B$ terms vanish at the input scale. In the pink shaded
region, the EWSB conditions cannot be satisfied, but in this case it is
because  $\mu^2 < 0$. Just to the right of the excluded region, we see
the equivalent of the focus-point strip, where the LSP is mostly
Higgsino. Still further to the right, we see two closely-spaced strips
corresponding to the funnel region with a mostly bino-like LSP. For this
choice of $\lambda$ and $\lambda^\prime$, the proton lifetime is
sufficiently long if $m_{1/2} \gtrsim 1.8$~TeV, but the Higgs mass is
$\lesssim 123$~GeV unless $m_{1/2} \gtrsim 2.7$~TeV. In the right panel
of Fig.~\ref{fig:twist13}, we again take all weights equal to 0, so that
$A_t = m_2$, $A_b = 0$, $A_\lambda = m_2$, $A_{\lambda'} = 0$, $B_H =
m_2$, and $B_\Sigma = 0$. In this case, the pink shaded region has
$m_A^2 < 0$ and we see the funnel strip running to values of $m_h >
125$~GeV. Comparing this with the left panel, we see the effect of the
non-zero value of $A_t$ on $m_h$. In both panels we see that points
along the dark matter strips with $m_{1/2} \gtrsim 1.7$~TeV have an
acceptably long proton lifetime.

Since we have both a focus point strip and a funnel region, there is more variation
in the computed elastic cross section. Corresponding to the left panel of Fig.~\ref{fig:twist13},
we considered points at $m_2 = 4000$ GeV with $m_{1/2} \simeq 2700$ GeV
(focus point with $m_\chi \simeq 900$ GeV) and $m_{1/2} \simeq 3200$ GeV
(funnel with $m_\chi \simeq 1160$ GeV). We found $\sigma^{\rm SI} \simeq (2.2 \pm 1.4) \times 10^{-8}$ pb
and $(1.2 \pm 0.7) \times 10^{-10}$ pb respectively. At higher $m_0 = 5000$ GeV, the cross section
on the focus point at $m_{1/2} \simeq 1065$ GeV drops to $(6.4 \pm 4.0) \times 10^{-9}$ pb
and on the funnel at $m_{1/2} \simeq 1530$ GeV drops to $(7.2 \pm 4.5) \times 10^{-11}$ pb.
When the weights are set to zero as in the right panel of  Fig.~\ref{fig:twist13},
we have only a funnel strip and the cross section is quite low.
For $(m_{1/2}, m_2)$ = (1920,3000), we find $\sigma^{\rm SI} \simeq (5.3 \pm 3.3) \times 10^{-11}$ pb
and for $(m_{1/2}, m_2)$ = (2965,4400), we find $\sigma^{\rm SI} = (2.2 \pm 1.4) \times 10^{-11}$ pb.

\section{Discussion}
\label{sec:discussion}

We have shown in this paper that, {\it if} the matter and Higgs supermultiplets
are all {\it untwisted}, super-GUT SU(5) models are unable to
provide simultaneously a long enough proton lifetime, a small enough
relic LSP density and an acceptable Higgs mass in the framework of
no-scale supergravity, even in the presence of a Giudice-Masiero term in
the K\"ahler potential. However, all of these phenomenological requirements
can be reconciled {\it if} one or both of the GUT Higgs fiveplets is
{\it twisted}. We have exhibited satisfactory solutions for various
values of the input super-GUT scale $M_{\rm in}$, the GUT Yukawa
couplings that are important in the RGEs above the GUT scale, and the
modular weights of the various matter and Higgs fields. All the examples
shown assume $\tan \beta = 7$: significantly smaller values of $\tan
\beta$ are largely excluded because $m_h$ is too small, and
significantly larger values of $\tan \beta$ are largely excluded because
the proton lifetime is too short. Spin-independent
dark matter scattering may be observable in some of the cases studied.

Although, as we have shown, many of the problems of the minimal SU(5)
GUT model may be resolved in the no-scale SU(5) super-GUT, including
rapid proton decay through dimension-5 operators, in a manner compatible
with the dark matter density and the Higgs mass, other issues such as
neutrino masses/oscillations remain unresolved. Moreover, the resolution
of the minimal supersymmetric SU(5) GUT problems within the super-GUT
and no-scale supergravity frameworks is quite constrained and
somewhat contrived. It also remains unclear how an SU(5) GUT model could be
embedded within string theory.

A natural alternative is the flipped SU(5)$\times$U(1) framework
proposed in~\cite{Barr,DKN,AEHN}, which resolves automatically the
problems mentioned above, and can be embedded with string theory.
Choosing even the simplest strict no-scale boundary conditions
$m_0=A_0=B_0 = 0$ at $M_{in}$ provides a very interesting flipped SU(5)
framework that satisfies all the constraints from present low-energy
phenomenology, including the relic dark matter density and the proton
lifetime, and makes interesting predictions for Run 2 of the
LHC~\cite{dvn}. Moreover, flipped SU(5) also contains
a rationale for $M_{in} > M_{GUT}$,  since the final unification of the
SU(5) and U(1) gauge couplings could well occur at the string scale.
We therefore plan to consider the possibility of a no-scale flipped
SU(5) super-GUT in a forthcoming paper.

\section*{Acknowledgements}

The work of J.E. was supported in part by the UK STFC via the research
grant ST/J002798/1. The work of D.V.N. was supported in part by the DOE
grant DE-FG02-13ER42020 and in part by the Alexander~S.~Onassis Public
Benefit Foundation. The work of N.N. and K.A.O. was supported in part by
DOE grant DE-SC0011842 at the University of Minnesota.




\end{document}